\documentclass[balance,upint,subscriptcorrection,varvw,mathalfa=cal=boondoxo,spanish,french,vietnamese,russian,greek,pdf-a,colorlinks]{asmeconf}

\hypersetup{%
	pdfauthor={John H. Lienhard},									  
	pdftitle={ASME Conference Paper LaTeX Template},                  
	pdfkeywords={ASME conference paper, LaTeX template, BibTeX style},
	pdfsubject = {Describes the asmeconf LaTeX template},			  
	pdflicenseurl={https://ctan.org/pkg/asmeconf},
}

\begin{document}


\ConfName{Proceedings of the OMAE 2024\linebreak 43rd International Conference on Ocean, Offshore \& Arctic Engineering}
\ConfAcronym{OMAE2024}
\ConfDate{June 9–14, 2024} 
\ConfCity{Singapore} 
\PaperNo{OMAE2024-125985}


\title{Flow-Induced Vibration of Flexible Hydrofoil in Cavitating Turbulent Flow } 
 
%
%
%

\SetAuthors{%
	Zhi Cheng\affil{1}\CorrespondingAuthor{vamoschengzhi@gmail.com}, 
	Rajeev Jaiman\affil{1}\CorrespondingAuthor{rjaiman@mech.ubc.ca}
	}

\SetAffiliation{1}{Department of Mechanical Engineering, The University of British Columbia, Vancouver, BC, V6T 1Z4, Canada. }


\maketitle

\versionfootnote{Documentation for \texttt{asmeconf.cls}: Version~\versionno, \today.}




\begin{abstract}

The flow-induced vibration and cavitation dynamics of three-dimensional flow past a cantilever flexible hydrofoil are investigated using a large eddy simulation (LES) model, a homogeneous mixture cavitation model and the structural modes superposition method.
The present work aims to explore a potential mechanism  responsible for a propeller singing behavior, and thus focuses on the synchronized hydroelastic coupling among the pressure pulsation inside the flow field, the cavitation generation and the structural vibration. 
To begin, we validate the tip vortex dynamics of a flexible hydrofoil against the available experimental. Our results demonstrate that the tip vortex shedding and the blade vibration are responsible for the intense peak in the low-frequency tonal components of the noise source, and the trailing-edge vortex shedding induces broadband components. 
Additionally,
the generation of sheet cavitation induces considerable synchronized hydrofoil vibration (subjected to a flutter-like response), and affects the pressure fluctuations in the flow field, which further dominate the features of the underwater noise sources.
It is suggested that the cavitation behavior and structural vibrations co-dominate the characteristics of singing noise from a propeller blade. 
\end{abstract}


%





\section{Introduction}
As a common underwater radiated noise (URN) phenomenon \citep{SMITH2022112863} in the marine industry, propeller singing behavior has always been a matter of interest in the industrial design and academic research \citep{Xincheng2023, Arndt2015, Maines1997, Higuchi1989}. Its generation and propagation pose safety and health concerns to the ship crew, passengers as well as marine mammals \citep{Savas2021, A-Man2023}. In addition, the potential causative factors underpinning it, flow-induced vibration (FIV) \citep{PARKINSON1989169, WilliamsonGovardhan2004}, could also lead to fatigue/fracture of vessel structures.
The most distinctive feature of the propeller singing behavior is that its noise spectrum is characterized by a narrow-band spectrum with a fairly high power level \citep{Arndt2015, Maines1997, Higuchi1989}. 
However, the dynamics of two potential issues closely correlated to propeller singing, i.e., the cavitation and the above-mentioned structural vibration of blades, are also complex. Therefore, the exact mechanism underlying propeller singing is still ambiguous.

As introduced above, propeller singing behavior exhibits tonal sound features \cite{Ross1989,carlton2018marine}, and specifically, the sound pressure level (SPL) of this peak is fairly (more than 15 dB higher) than the neighboring spectral level, which could be heard in the aft cabin of the vessel. 
Furthermore, the propeller singing behavior will appear at a certain speed range, not overall range, for the rotating propeller. However, once singing occurs, the correlated sound will be sustained for this speed range, and the tonal frequency does not change with the propeller rotating speed. 
With respect to the associated investigation, in addition to some theoretical analysis \citep{Bosschers2009InvestigationOT,thomson_1880, Bosschers2009} as well as experimental works \citep{pennings_2015,ye_wang_shao_2023,Higuchi1989}, high-fidelity calculations have been applied recently to predict and analyze the URN problem and also the accompanied vortex-shedding, cavitation, and acoustics issues.
Compared to analyzing the overall noise features for a full-scale propeller \citep{posa_broglia_felli_cianferra_armenio_2022, jmse8030158}, the more detailed focus on the dynamics of separate blade/hydrofoil assists more in the exploration of the mechanism underpinning URN problems, especially the behavior of propeller singing.

The noise generation of underwater hydrofoils is closely associated with cavitation generation \citep{JiBin2023}. 
Yang {\it et al.} \cite{YangChen2022} studied the unsteady cavitation and non-cavitation flow of the three-dimensional (3D) rigid hydrofoil, with multiphase flow solved using a large eddy simulation (LES) model coupled with the modified Schnerr-Sauer cavitation model.  Ffowcs Williams-Hawkings acoustic method is applied for the acoustic calculation. 
This study stated that the cavitation transformation and pressure fluctuations are responsible for the low-frequency noise source generated on the body surface. The work of Wang {\it et al.} \cite{Xincheng2023} focused on the potential relationship between URN features and tip vortex cavities. A new numerical prediction method is introduced to capture the theoretical dispersion relations of the surface waves traveling on tip vortex cavities. Volume fluctuation of the cavity is found to dominate the low- and medium-frequency range of the URN spectrum.
However, the detailed intrinsic correlation between cavitation and singing behavior, i.e. the production of tonal noise, remains not fully understood.

There are some works \citep{blake2017mechanics,Fischer2008} claiming that fluid-structure interactions are responsible for the production of high-pitched sound (i.e., propeller singing). It is also proposed that the behavior occurs due to the matching of the main vortex shedding frequency with the blade's structural natural frequency. This argument is in conformance with the above-mentioned features that singing behaviors are sustained in a certain rotor speed range with constant tonal frequency.
In this case, FIV has been suggested to be a potential underlying factor resulting in propeller singing behavior \citep{blake2017mechanics,Fischer2008}, although other research work partially refuted this statement \citep{Arndt2015}. 
With respect to the structural response of hydrofoil, Chae {\it et al.} \cite{EunJung2016} investigated the FIV response of the flexible hydrofoil via numerical and experimental works and focused on the effect of the inflow velocity, angle of attack, and added mass effects.
The induced vibration response also has an interactive relationship with the cavitation generated on the hydrofoil surface or tip. Sun {\it et al.} \cite{Tiezhi2021} investigated the cavitation process of hydrofoils with structure response calculated using the finite element method and flow field solved using large eddy simulation and Schnerr–Sauer model. The responses, including structural displacements, cavitation, and pressure fluctuations, are compared between rigid and flexible hydrofoils. Sun {\it et al.} \cite{Tiezhi2021} concluded that the flexible hydrofoil exhibits smaller and more diffuse areas of cavitation compared with the rigid hydrofoil.
Dong {\it et al.} \cite{Songwen2023} further explored both the flow dynamic and acoustic response of flexible hydrofoils under cavitation-induced vibration and different structural materials compared. This work also noted that material optimization is able to suppress cavitation while reducing turbulence intensity along with the attenuation of noise generation and propagation.

However, earlier works have not examined the issue of the potential relationship between noise source and the three-dimensional (3D) vibration response of a flexible blade tip in the presence of turbulent cavitating flow.  
The closest research is the above-introduced work conducted by Wang {\it et al.} \cite{Xincheng2023}, but the concerned tip hydrofoil is in rigid form. It is known that the hydrofoil shape of a blade tip is tapering and is more susceptible to deformation as well as vibration than a hydrofoil with a chord length constant along the spanwise direction. In addition, cavitation is more likely to occur at the tip area due to a relatively larger Reynolds number. Therefore, the FIV study of the flexible tip hydrofoil is of great significance in analyzing the underlying mechanism of propeller singing.
Based on the above introduction, present work will focus on the explosion of the detailed characterization of 3D FIV response of flexible tip hydrofoil considering the effect of cavitation processing.

The paper is structured as follows:
Section 2 details the numerical and analytical methods listed above.
The accuracy of the implemented models used herein is also validated in Section~3.
In Section 4, the FIV response and potential singing behavior for the hydrofoil are analyzed. A systematic comparison is conducted between rigid and flexible hydrofoils regarding the flow and structural responses and the impact of cavitation is explored.
In Section 5, the key results of this study are summarized.


\section{Numerical methodology}

\subsection{Fluid-structure interaction}
With respect to the concerned configuration consisting of a flexible (elastic) cantilevered hydrofoil submerged in a 3-D uniform water flow in this work, the computational fluid dynamics/computational structure dynamics method is applied to calculate the system response. 
The concerned fluid dynamics consists of liquid and vapor phases and is assumed to exist as a continuous homogeneous mixture. The phase indicator $\alpha _l$($\boldsymbol{x}$,$t$) represents the phase fraction of the liquid phase in the fluid mixture. Where $\boldsymbol{x}$ and $t$ are spatial and temporal coordinates. The density $\rho$ and dynamic viscosity $\mu$ of the fluid are considered as a weighted combination of the liquid and vapor phases:

	\begin{equation}\label{}
\rho =\rho _l\alpha _l+\rho _v\left( 1-\alpha _l \right) ,
	\end{equation}
	\begin{equation}\label{}
\mu =\mu _l\alpha _l+\mu _v\left( 1-\alpha _l \right),
	\end{equation}
where $\rho_l$ and $\rho_v$ are the densities of the pure liquid and vapor phases, respectively, and $\mu_l$ and $\mu_v$ are the dynamic viscosity of the liquid and vapor phases.

The governing equations of the flow dynamics are the unsteady incompressible Navier-Stokes (N-S) equations combined with the Schnerr-Sauer cavitation model, while the boundary changes induced by the body motion are resolved based on an arbitrary Lagrangian-Eulerian (ALE) scheme. The N-S equations in the ALE scheme are expressed as:
	\begin{equation}\label{eq:ns1}
\rho \frac{\partial \boldsymbol{u}}{\partial t}+\rho \left( \boldsymbol{u}-\boldsymbol{\hat{u}} \right) \cdot \nabla \boldsymbol{u}-\nabla \cdot \boldsymbol{\sigma }=\boldsymbol{f}_f,\,\,   \text{on}\left( \boldsymbol{x},t \right) \in \varOmega 
	\end{equation}
	\begin{equation}\label{eq:ns2}
\frac{\partial \rho}{\partial t}+\rho \nabla \cdot \boldsymbol{u}+\left( \boldsymbol{u}-\boldsymbol{\hat{u}} \right) \cdot \nabla \rho =0,\,\,   \text{on}\left( \boldsymbol{x},t \right) \in \varOmega 
	\end{equation}
	\begin{equation}\label{}
\nabla \cdot \boldsymbol{u}=\left( \frac{1}{\rho _l}-\frac{1}{\rho _l} \right) \left( \dot{m}_c+\dot{m}_v \right) 
	\end{equation}
where $\boldsymbol{u}$ is the fluid flow velocity, $\boldsymbol{\hat{u}}$ is the velocity component of mesh motion, $\boldsymbol{f}_f$ is the body force applied to the fluid, 
$\boldsymbol{x}$, i.e., $(x,y,z)$, is the Cartesian coordinates in the fluid domain $\varOmega$.
$\dot{m}_c$ and $\dot{m}_v$ are the mass transfers in the condensation and vaporization processes respectively.
$\boldsymbol{\sigma}$ could be expressed as:
	\begin{equation}\label{}
\boldsymbol{\sigma }=\boldsymbol{\sigma }^n+\boldsymbol{\sigma }^{sgs},
\end{equation}
where $\boldsymbol{\sigma }^n$ and $\boldsymbol{\sigma }^{sgs}$ are the Cauchy stress tensor for a Newtonian fluid and the subgrid-scale turbulent stress tensor, respectively.
$\boldsymbol{\sigma }^n=-p \boldsymbol{I}+\mu \left( \nabla \boldsymbol{u}+\left( \nabla \boldsymbol{u} \right) ^T \right)$,
where $p$ represents the fluid pressure. 
$\boldsymbol{\sigma }^{sgs}$ is modeled using a spatial filtering large eddy simulation (LES) model \cite{Germano1991, Gatski_Speziale_1993}.

\subsection{Cavitation model}

As the solution of the scalar transport equation,  indicator $\alpha_l$ is obtained as the liquid volume fraction. The scalar transport equation could also be written in the ALE framework as:
	\begin{equation}\label{}
    \frac{\partial \alpha _l}{\partial t}+\left( \boldsymbol{u}-\boldsymbol{\hat{u}} \right) \cdot \nabla \alpha _l=\frac{\rho}{\rho _l\rho _v}\left( \dot{m}_c+\dot{m}_v \right),\,\,\text{on} \left( \boldsymbol{x},t \right) \in \varOmega  
	\end{equation}
The Schnerr-Sauer cavitation model \citep{sauer2001development} (based on the transport equations) mathematically expresses the vaporization and condensation process and is able to capture the cavitation phenomenon more realistically.
The Schnerr-Sauer cavitation model used in this paper is implemented in our in-house code and expressed as:
	\begin{equation}\label{}
\dot{m}_c=\frac{3\rho _l\rho _v}{\rho R_B}\sqrt{\frac{2}{3\rho _l\lvert p-p_v \rvert}}\cdot C_c\alpha _l\left( 1-\alpha _l \right) \max \left( p-p_v,0 \right) 
	\end{equation}

	\begin{equation}\label{}
\dot{m}_v=\frac{3\rho _l\rho _v}{\rho R_B}\sqrt{\frac{2}{3\rho _l\lvert p-p_v \rvert}}\cdot C_v\alpha _l\left( 1+\alpha _{nuc}-\alpha _l \right) \min \left( p-p_v,0 \right) 
	\end{equation}

	\begin{equation}\label{}
R_B=\left( \frac{3}{4\pi n_0}\frac{1+\alpha _{nuc}-\alpha _l}{\alpha _l} \right) ^{1/3},
	\end{equation}

	\begin{equation}\label{}
\alpha _{nuc}=\left( \frac{\pi n_0d_{nuc}^{3}}{6} \right) /\left( 1+\frac{\pi n_0d_{nuc}^{3}}{6} \right).
	\end{equation}
where $C_c$ and $C_v$ are the condensation and vaporization model coefficients, respectively (equal to 1$\times$$10^{-3}$ and 5$\times$$10^{-3}$, respectively, herein); 
$p_v$ is the liquid saturation pressure; 
$R_B$ is the equivalent radius of the vapor volume;
$d_{nuc}$ is the local nuclei diameter (whose value is equal to 2.5$\times$$10^{-6}$ herein); and, $n_0$ is the number of bubbles per unit volume (whose value is equal to 1.0$\times$$10^{13}$ herein).
Detailed information on the model and properties determination refers to the works of Kashyap \& Jaiman \cite{KASHYAP2023104276} and Sauer \& Schnerr \cite{sauer2001development}.

\subsection{Structural dynamics for flexible  blade}

\begin{figure}
    \centering
    \includegraphics[width=1.0\linewidth]{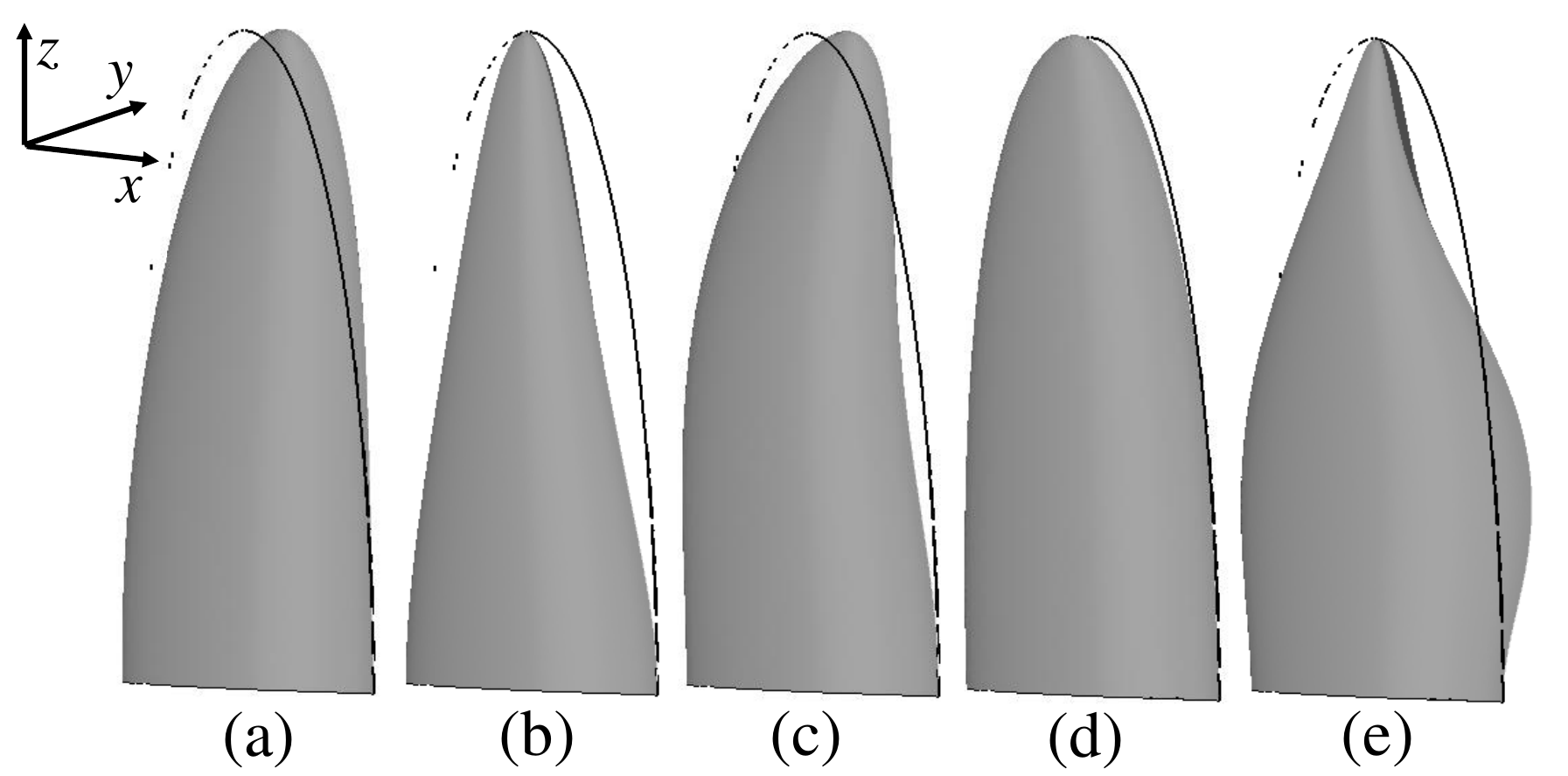}
	\caption{First five mode shapes of a flexible cantilever hydrofoil. The fourth mode in panel (d) vibrates in $x-$direction.}
	\label{5modeshapes}
 \end{figure}

The flexible cantilever hydrofoil concerned here is regarded as a structure with relatively small three-dimensional deformations, and the present hydrofoil is modeled using eigenmodes obtained from solving the eigenvalue problem
derived from the linear elastic structural model.
We solve the structural displacements $\boldsymbol{\omega }^s(\boldsymbol{x}^s,t)$ at co-ordinates $\boldsymbol{x}^s$ using a linear elastic control equation excited by the distributed unsteady force $f^s$ (acting on per unit volume). The motion of the flexible cantilever hydrofoil is governed by the fluid forces and involves integrating pressure and shear stress effects on the blade surface. Neglecting the damping and shear effects, the control equations for the flexible cantilever structure are expressed as:
\begin{equation}\label{eq:sm1}
\boldsymbol{\rho }^s\frac{\partial ^2\boldsymbol{\omega }^s}{\partial t^2}=\nabla \cdot \boldsymbol{\sigma }^s+\rho ^sf^s,
	\end{equation}
where $\rho^s$ is the solid density, $\boldsymbol{\sigma }^s$ is the Cauchy stress tensor of the solid, which is defined as (for the linear elastic material):
\begin{equation}\label{eq:sm2}
\boldsymbol{\sigma} ^s=\lambda ^str\left( \boldsymbol{\epsilon} ^s \right) +2\mu ^s\boldsymbol{\epsilon} ^s,
	\end{equation}
where $\lambda ^s$ and $\mu ^s$ are the Lame’s parameters and $\boldsymbol{\epsilon}$ is the infinitesimal strain tensor, which is defined as:
\begin{equation}\label{eq:sm3}
\boldsymbol{\epsilon }^s=\frac{1}{2}\left( \nabla \boldsymbol{\omega}^s+\left( \nabla \boldsymbol{\omega}^s \right) ^T \right).
	\end{equation}
To obtain the eigenvalue problem, we separate the spatial and temporal components of the displacement response as:
\begin{equation}\label{eq:sm4}
\boldsymbol{\omega }^s\left( \boldsymbol{x}^s,t \right) =\sum_{n=1}^{\infty}{\boldsymbol{\hat{\omega}}_{n}^{s}\left( \boldsymbol{x}^s \right)}e^{iw_nt},
\end{equation}
where $\boldsymbol{\hat{\omega}}_{n}^{s}$ denotes an eigenvector or mode shape of the structure, and $w_n$ represents the corresponding eigenvalue or modal frequency. Substituting Eq. \ref{eq:sm4} in Eq. \ref{eq:sm1} and neglecting body forces, we obtain the
following representation:
\begin{equation}\label{eq:sm5}
-\rho ^sw_{n}^{2}\boldsymbol{\hat{\omega}}_{n}^{s}e^{iw_nt}=\nabla \cdot \boldsymbol{\sigma }^s\left( \boldsymbol{\hat{\omega}}^s \right) e^{iw_nt},
\end{equation}
which owing to linear elastic modeling can be simplified to:
\begin{equation}\label{eq:sm6}
\left( \mathcal{L}+\rho ^sw_{n}^{2}\boldsymbol{I} \right) \boldsymbol{\hat{\omega}}_{n}^{s}=0,
\end{equation}
where $\mathcal{L}$ denotes the linear differential operator associated with Eq. \ref{eq:sm3} and Eq. \ref{eq:sm1}, completing the definition of the structural eigenvalue problem. The boundary conditions of the cantilevered arrangement are given by:
\begin{equation}\label{eq:sm7.1}
 \boldsymbol{\omega }^s\left( \boldsymbol{x}^s,t \right) \rvert_{z=0}=0,
\end{equation}
\begin{equation}\label{eq:sm7.2}
 \frac{\partial \boldsymbol{\omega }^s\left( \boldsymbol{x}^s,t \right)}{\partial z} \rvert_{z=0}=0,
\end{equation}
\begin{equation}\label{eq:sm7.3}
 \boldsymbol{\sigma }^{\boldsymbol{s}}\boldsymbol{n}^{\boldsymbol{s}} \rvert_{z\ne 0}=0.
\end{equation}
The solution to Eq. \ref{eq:sm6} is obtained through a finite element discretization of the structural domain. As introduced above, a mode superposition method is considered for resolving Eq. \ref{eq:sm1} and characterizing the dynamic response of the hydrofoil \citep{Heydari2022}.
Calculix \cite{Calculix} software is applied herein to extract the shape of each mode and the corresponding structural natural frequency.
First-ten dominating structural modes are considered in the present work.
The structural properties of present flexible cantilever hydrofoil are defined with Young's Module $E$ of 6 $\times$ $10^{8}$ $Pa$, Poisson's ratio $\nu$ of 0.33, and mass ratio $m^*$ of 7.65, resulting in the corresponding eigenvalues and nature frequencies (for each mode) displayed in Table \ref{mode_shapes_f}. The diagram of first-five mode shapes is shown in Fig. \ref{5modeshapes}.

\begin{center}
\begin{table}
\caption{Eigenvalue $k_i$ and structural nature frequency $f_{n,i}$ of each mode for cantilever flexible hydrofoil concerned in present work.}
\centering
\begin{tabular}{  c  c  c  c  c  c  c  c  c  c  }
\hline
Mode shape &  $k_i$ (N/m)  & $f_{n,i}$ (Hz) \\
\hline
1             & 0.2698 $\times$ $10^{6}$    & 82.7  \\
2             & 0.2690 $\times$ $10^{7}$    & 261.0  \\
3             & 0.4354 $\times$ $10^{7}$    & 332.1  \\
4             & 0.5506 $\times$ $10^{7}$    & 373.5  \\
5             & 0.1492 $\times$ $10^{7}$    & 614.7   \\
6             & 0.9849 $\times$ $10^{8}$    & 706.4  \\
7             & 0.3496 $\times$ $10^{8}$    & 941.0  \\
8             & 0.3620 $\times$ $10^{8}$    & 957.6  \\
9             & 0.3840 $\times$ $10^{8}$    & 986.2  \\
10            & 0.4034 $\times$ $10^{8}$    & 1010.9  \\
\hline
\end{tabular}
\label{mode_shapes_f}
\end{table}
\end{center}

A standard Lagrangian finite element technique is applied to adapt the mesh to the new geometry of the domain.
A nonlinear partitioned iterative approach
based on the nonlinear iterative force correction (NIFC) scheme is applied to solve the fluid-structure interaction that links the fluid flow equations (\ref{eq:ns1}, \ref{eq:ns2}) with the structural equation of motion (\ref{eq:sm1}).
The approach for integrating the structural equations of motion, treating the fluid-structure interface, and solving ALE mesh motion refers to the works and introduction of Darbhamulla \& Jaiman \cite{darbhamulla2023finite}, Kashyap \& Jaiman \cite{KASHYAP2023104276} and Heydari {\it et al.} \cite{Heydari2022}.

\begin{figure}
	\centering
	\begin{subfigure}[b]{1.0\linewidth}
		\includegraphics[width=\linewidth]{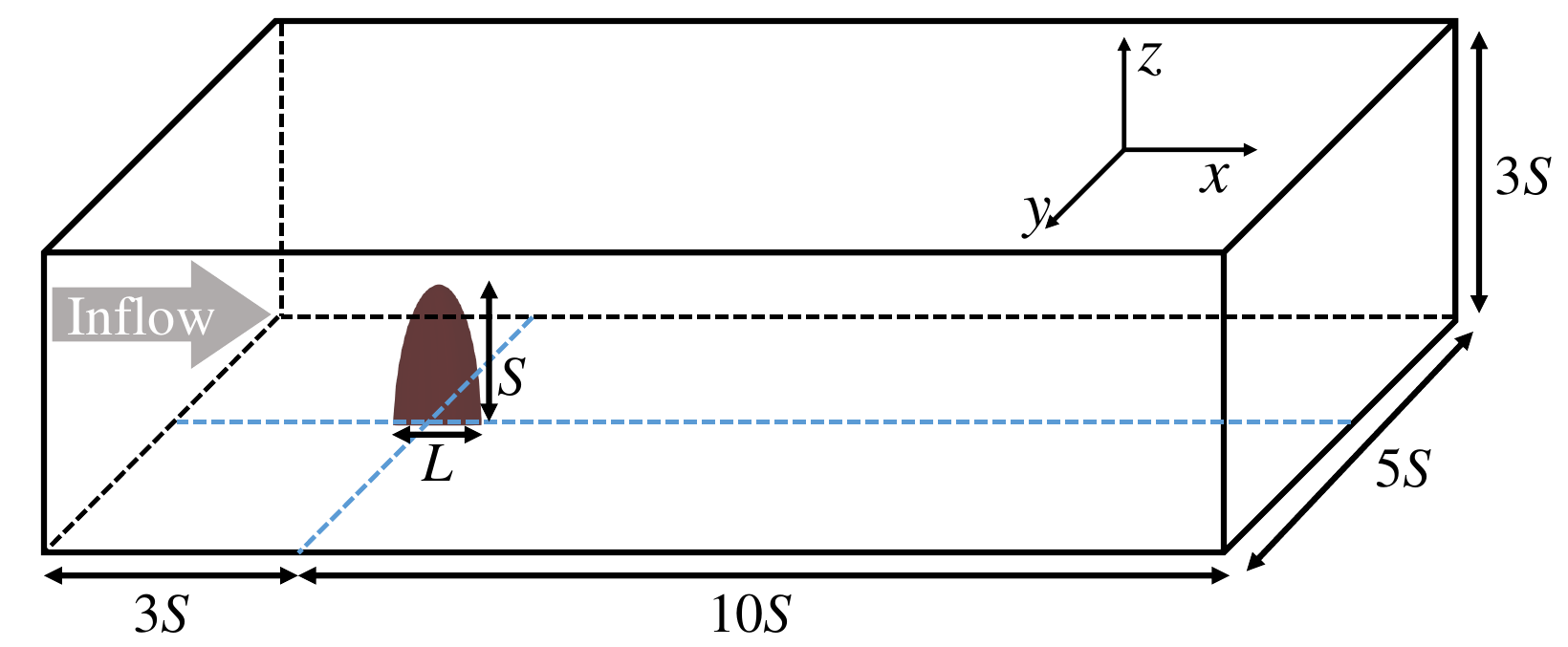}
	\caption{}
	\label{compu_domain}
	\end{subfigure}
	\begin{subfigure}[b]{0.7\linewidth}
		\includegraphics[width=\linewidth]{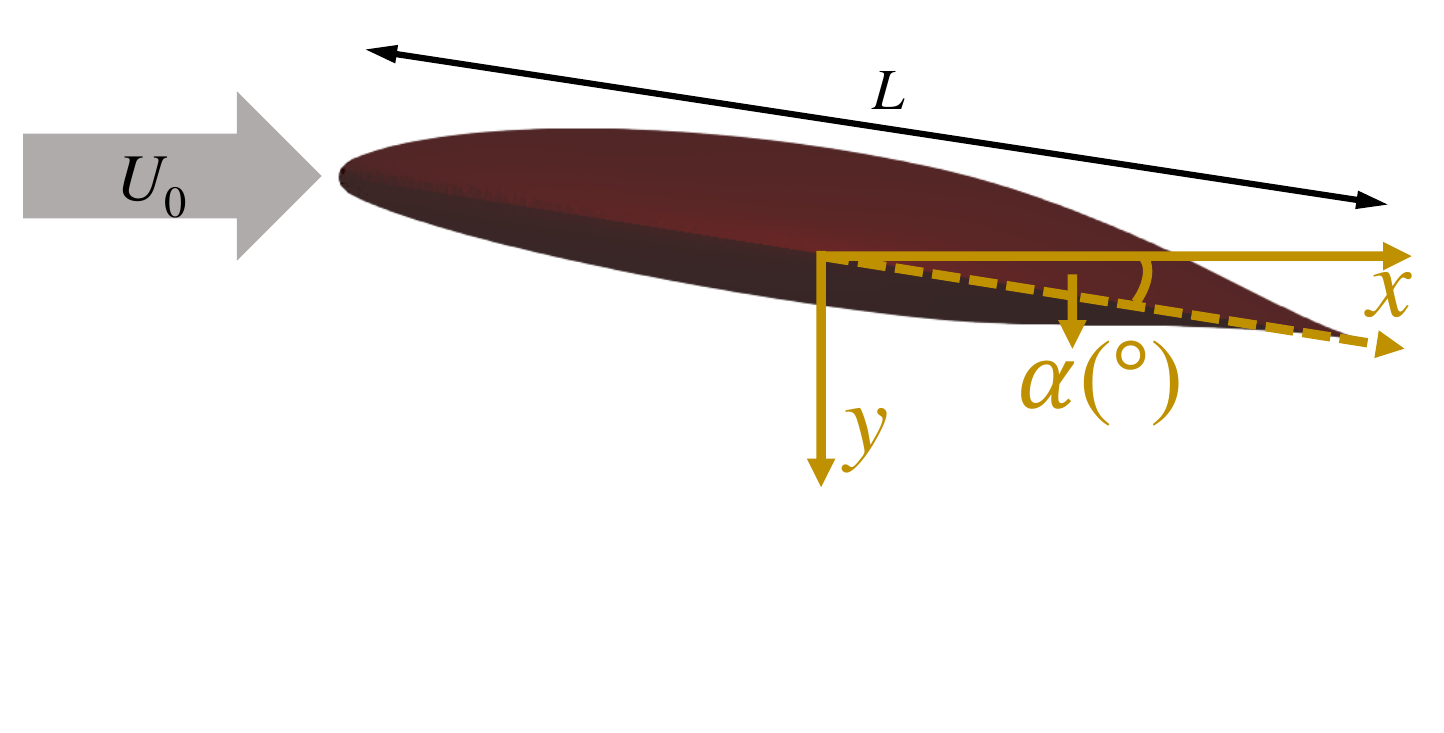}
	\caption{}
	\label{blade_diagram}
	\end{subfigure}
	\caption{Flexible hydrofoil set-up: (a) overall computational domain, and (b) top-view of present cantilever hydrofoil.}
	\label{diagram}
\end{figure}

\begin{figure}
    \centering\includegraphics[width=1.0\linewidth]{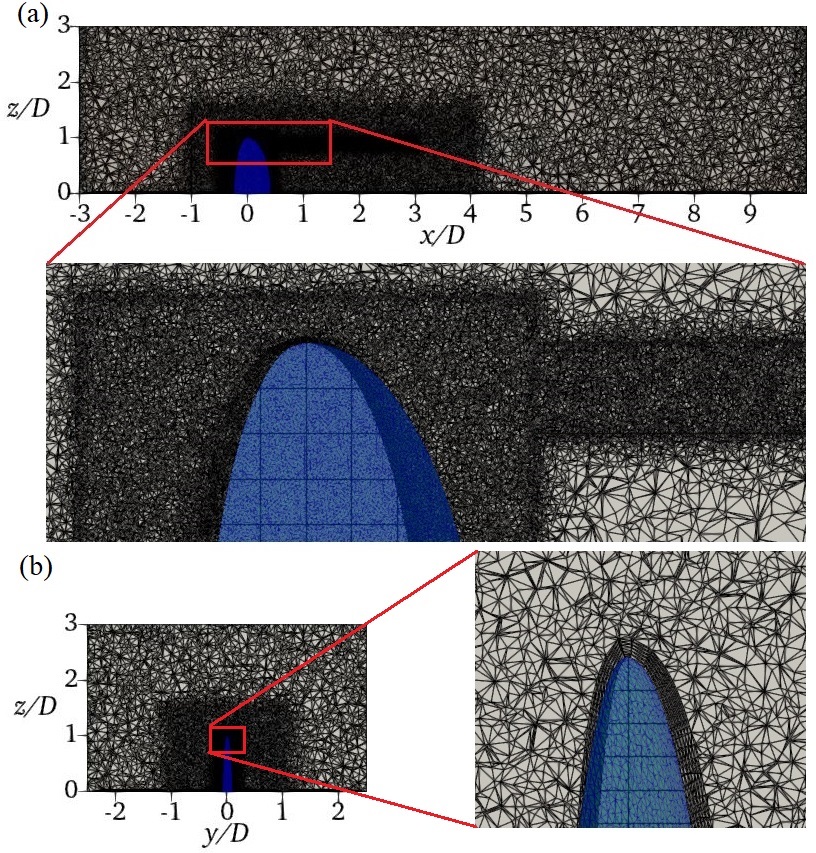}
	\caption{Overall mesh and the expanded view of the immediate vicinity of the body surface in (a) $x-z$ and (b) $y-z$ planes.}
	\label{mesh3}
 \end{figure}

\section{Computational domain and model validation}
\label{Secdomain}

Figures \ref{diagram}a and \ref{diagram}b exhibit an overall diagram and a top view of the present configuration of flexible NACA 66$_2$-415 elliptic hydrofoil (in cantilever pattern with bottom-face fixed) submerged in uniform water inflow.  The angle of attack ($\alpha$), span ($S$), and maximum chord ($L$) of the concerned hydrofoil were 9$^{\circ}$, 83.6 mm and 100.0 mm, respectively. 
With respect to the present three-dimensional (3-D) computational domain in Figure \ref{diagram}a, the position of the hydrofoil bottom surface is at the centerline of transverse (or, cross-stream) direction (i.e., $y=$ 0), and situated at 3$S$ downstream from the inlet boundary in $x-$direction. 
A Dirichlet boundary condition was prescribed for the incident flow velocity $\boldsymbol{u} = (U_0, 0, 0)$ on the inlet face (i.e., single upstream patch) in Fig. \ref{diagram}a, with $U_0$ fixed as 10 m/s in present work.
A Neumann boundary condition is imposed on the velocity at the outflow (outlet) boundaries, i.e., the single downstream patch of the domain, 
A non-slip boundary condition is applied on the bottom patch.
symmetrical boundary conditions are applied to two-sided and up patches to avoid blockage effects. The cavitation number is defined as
$\vartheta _n=(p-p_{v}) / (0.5\rho_l (U_0) ^2)$, where $p_v$ is saturation pressure.

\begin{center}
\begin{table}
\centering
\caption{Surface force coefficients for different mesh conditions with rigid hydrofoil.}
\begin{tabular}{  c  c  c  c  c  c  c  c  c  c  }
\hline
&  Number of elements & $C_{x,rms}$ & $C_{y,rms}$  & $C_{z,rms}$  \\
\hline
 Mesh 1 & 2,117,540 & 0.0552 & 0.4217 & 0.0295 \\ 
 Mesh 2 & 4,475,608 & 0.0584 & 0.4621 & 0.0312 \\ 
 Mesh 3 & 6,083,799 & 0.0608 & 0.4978 & 0.0338 \\ 
 Mesh 4 & 7,987,184 & 0.0612 & 0.4981 & 0.0340\\ 
\hline
\end{tabular}
\label{thrust_com}
\end{table}
\end{center}

\begin{figure}
    \centering\includegraphics[width=1.0\linewidth]{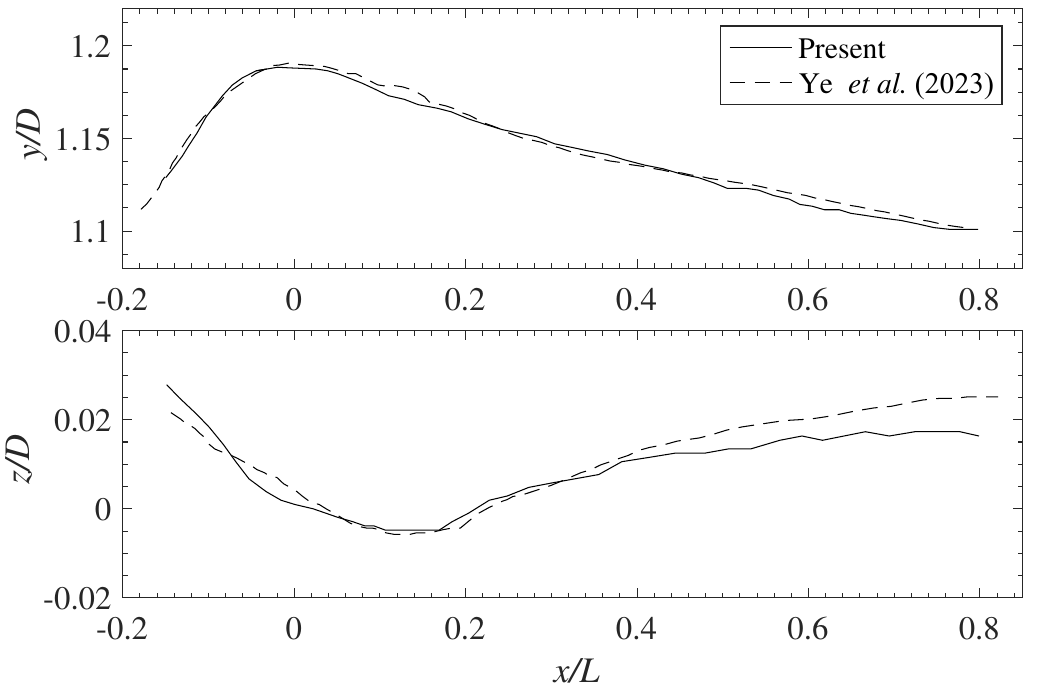}
	\caption{Comparison between present work and Ye {\it et al.} \citep{ye_wang_shao_2023} for the trajectory of the time-averaged tip vortex.}
	\label{trajectary_com}
 \end{figure}

\section{Results and Discussion}
\label{SecDiscussion}
The mesh dependency discussion is conducted via the simulation of the present NACA 66$_2$-415 elliptic hydrofoil in rigid/solid form.
The corresponding drag and lift coefficients ($C_x$ and $C_y$) at different mesh qualities are calculated and the associated results are summarized in Table \ref{thrust_com}.
It could be observed that the relative differences of each parameter between mesh 1 to mesh 2 are considerable, but all decrease to a value smaller than 0.5\% as the mesh is refined to mesh 3 (fine) and mesh 4 (very fine).
As a consequence, the strategy applied by mesh 3 is adopted in all the configurations of the present work to achieve the best balance of calculation time and accuracy.
To follow up, we use mesh 3 to calculate the present configuration and compare the trajectory of the time-averaged tip vortex between present results and other accessible experimental work \citep{ye_wang_shao_2023}.  The results summarized in Fig. \ref{trajectary_com} indicate high conformance between this study and other results, which indicates the correctness of the present model.
The fluid-solid interaction model for flexible structures and the fluid model involving cavitation phenomena applied in this study are implemented in the in-house  FSI solver, whose detailed description and model validation refer to our previous works of   Kashyap \& Jaiman \cite{KASHYAP2023104276, Kashyap202119} and Heydari {\it et al.} \cite{Heydari2022}.
Figure ~\ref{mesh3} shows the overall and extended view of the computational domain corresponding to mesh 3.

\begin{figure}
\centering
\centering\includegraphics[width=1.0\linewidth]{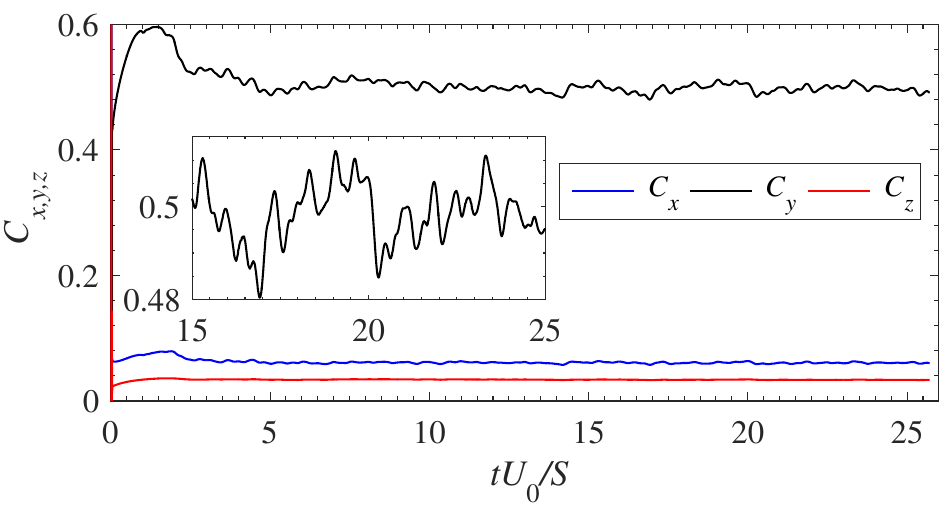}
\caption{Time-histories of lift coefficients in $x-$, $y-$, and $z-$directions for rigid hydrofoil without cavitation.}
\label{Cxyz_timehistory}
\end{figure}

\subsection{Rigid blade without cavitation}

We first explore the configuration consisting of concerned hydrofoil in rigid form. 
The time-histories of surface force coefficients in $x-$, $y-$, and $z-$ are displayed in Fig. \ref{Cxyz_timehistory}. It could be observed that the mean value as well as the fluctuating components of the pressure variation in $y-$ and $z-$directions are maximum and minimum, respectively. Therefore, the following contents will focus on the response of the system dynamics in the $x-$ and $y-$directions.
Torque force is not displayed here owing that rotational displacement underlying mode 5 do not has a dominate impact on structural response herein.

\begin{figure}
\centering
\centering\includegraphics[width=0.8\linewidth]{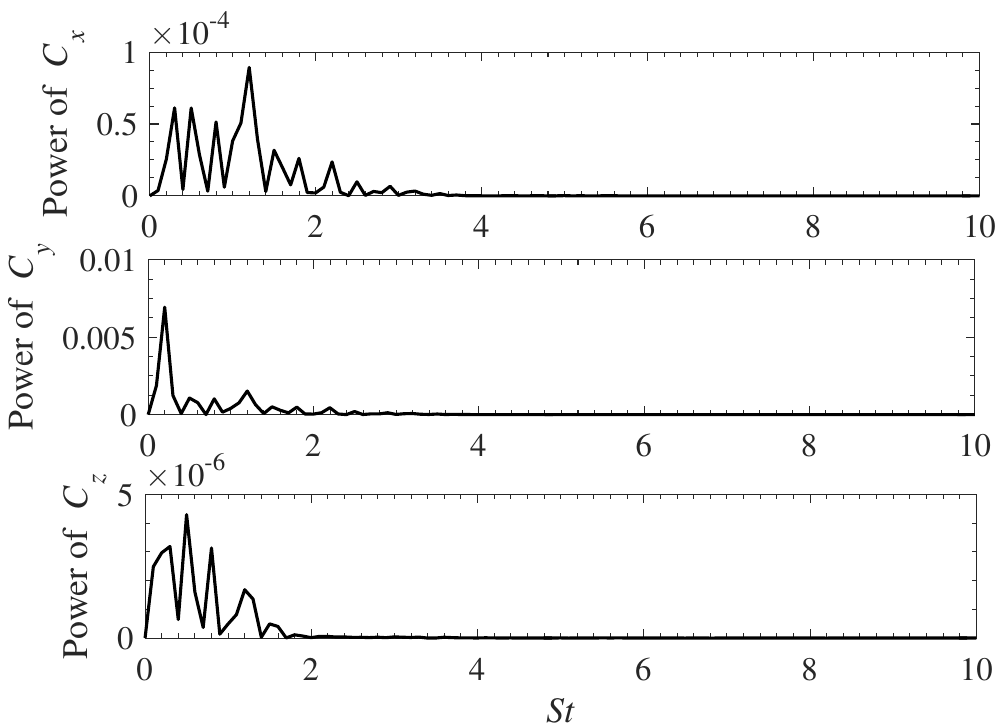}
\caption{Spectrum of lift coefficients in $x-$, $y-$, and $z-$directions for rigid hydrofoil without cavitation.}
\label{Cxyz_fft}
\end{figure}

\begin{figure}
    \centering\includegraphics[width=0.9\linewidth]{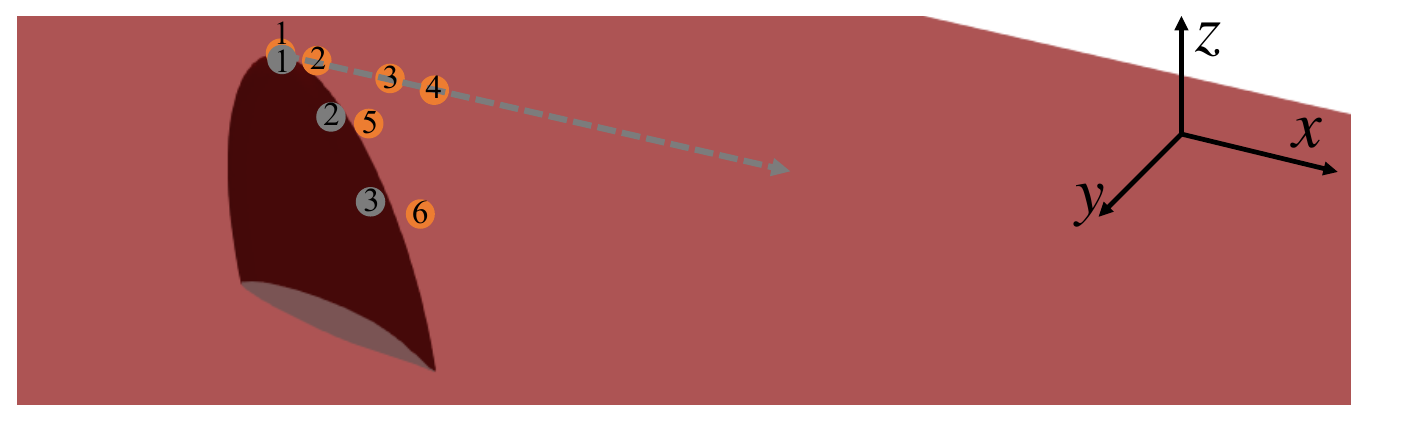}
	\caption{Diagram of the monitor locations.}
	\label{monitors}
\end{figure}

\begin{figure}
\centering
\centering\includegraphics[width=0.85\linewidth]{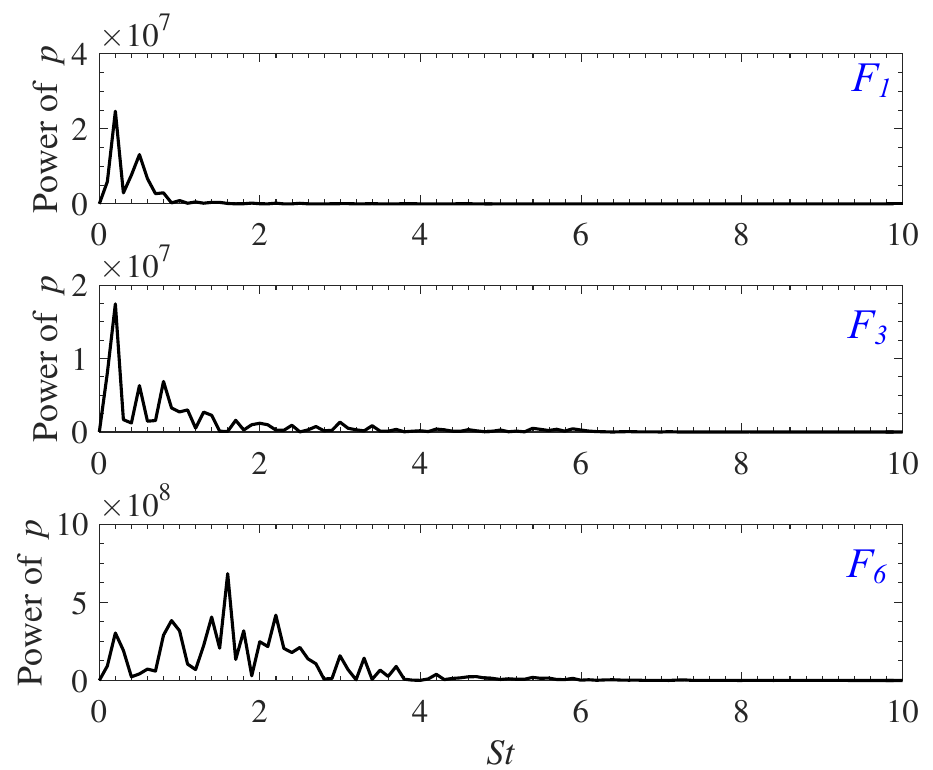}
\caption{Spectrum of pressure fluctuation at monitors $F_1$, $F_3$, and $F_6$ for rigid hydrofoil without cavitation.}
\label{field_p_fft}
\end{figure}

The fluctuation components inside $C_{x,y,z}$ variations are extracted and the corresponding spectrums (as a function of $St$) are shown in Figure \ref{Cxyz_fft}. $St$ (=$fS/U_0$) is the Strouhal number, where $f$ is frequency.
With respect to $C_y$ variation, the most intense peak is located at $St$ of 0.2 (i.e., $f$ = 20 Hz).
For the streamwise force $C_x$, besides the peak with $St$ close to 0.2, there are other multi-components ranging from $St$ of 0.4 to about 2.0, in which the strongest peak corresponds to $St$ = 1.5. A similar situation appears for the $C_z$.
Those frequency components are correlated to the structural response of the flexible hydrofoil and also pressure fluctuations underpinning propeller singing introduced later in the present work. In this case, we are curious about which dynamical structures each of the above frequency components correspond to.

To answer the above question, a specific attention is placed on the pressure variations at certain points inside fluid domain and on hydrofoil surface.
Figure \ref{monitors} indicates the locations of the monitors, in which the grey-color and yellow-color points represent the monitors on the surface and inside the flow domain, respectively. Field monitors $F_{1,2,3,4}$ locate downstream of the blade tip with the same vertical ($z-$) height.
The spectrum for the pressure fluctuation at yellow-marked field monitors 1, 3, and 6 (i.e., $F_{1,3,6}$) are shown in Fig. \ref{field_p_fft}.
It should be noted that monitor $F_1$ is on the tip of the hydrofoil, $F_3$ is located in the trajectory of shedding tip vortex with $x/S$ = 0.5, and $F_6$ is very close to the trailing edge of the hydrofoil with $z/S$ = 0.4.

The first as well as the second panel of Fig. \ref{field_p_fft} indicate that the dominant frequency of the tip vortex-shedding behavior is located at $St$ = 0.2, which is consistent with the intense frequency of the second panel in Fig. \ref{Cxyz_fft}. This indicates that the tip vortex-shedding mechanism underpins the lift variation in the $y-$direction when the hydrofoil is at rigid and stationary.
On the other hand, the third panel in Fig. \ref{field_p_fft} exhibits a clear broadband feature, with correlated $St$ varying from 0 to 4. It is expected that these broadband components are associated with the continuous trailing-edge vortex-shedding structure on the blade surface in the spanwise direction. 
Furthermore, the chord length of the blade continues to get smaller along the spreading direction, and this tapering geometrical feature is one of the factors for the continuous change in the surface vortex-shedding frequency and also the broadband features of the pressure spectrum.
Those pieces of information suggest that vortex shedding at the trailing edge and tip of the blade (at rest status) will affect the broadband and tonal components of its hydrodynamic noise, respectively.

\begin{figure}
\centering
\centering\includegraphics[width=1.0\linewidth]{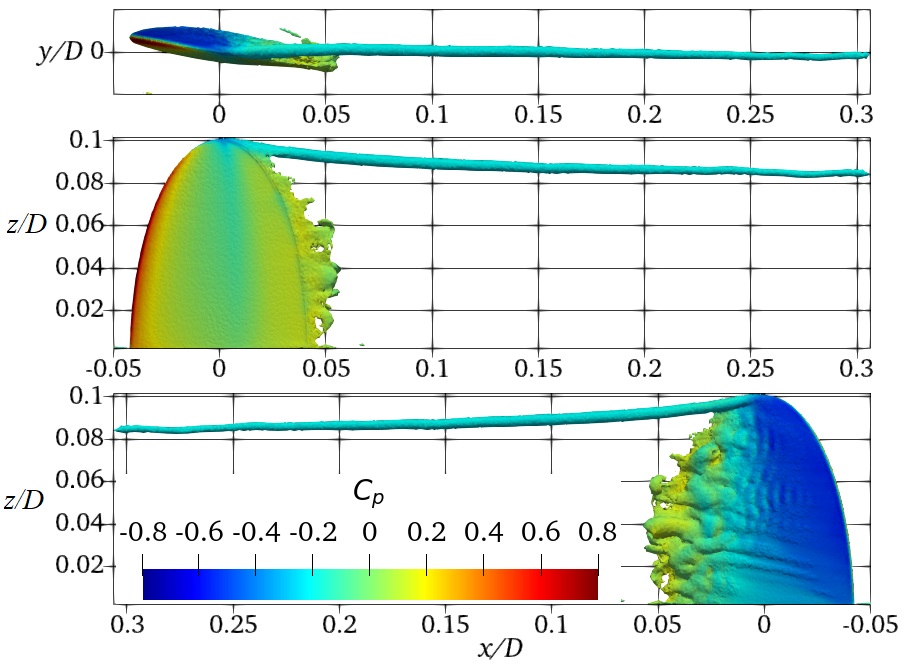}
\caption{Instantaneous isopleths of the vorticity from different views for rigid hydrofoil without cavitation.}
\label{noca_fixed_vorticity}
\end{figure}

\begin{figure}[b]
\centering
\centering\includegraphics[width=0.9\linewidth]{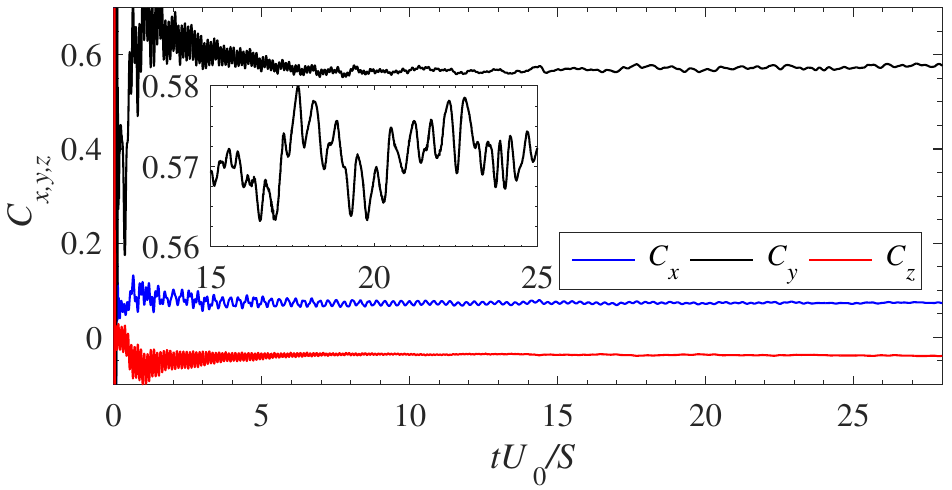}
\caption{Time-histories of lift coefficients in $x-$, $y-$, and $z-$directions for flexible hydrofoil without cavitation.}
\label{noca_9degU10_0.2E_Cxyz_timehistory}
\end{figure}

Figure \ref{noca_fixed_vorticity} shows the contour of the instantaneous vorticity magnitude in the $x-y$ and two $x-z$ orthogonal planes (from front and back views) for flow past the rigid hydrofoil, with the distribution of pressure coefficients depicted on it.
At the front and back of the hydrofoil are the pressure side and suction side, respectively, where the suction side will generate cavitation later in the following cavitating calculation due to the low-pressure distribution.
Additionaly, in addition to the clear vortex shedding trajectory behind the top of the hydrofoil tip, the suction side also showed the behavior of fluid transitioning from laminar flow into turbulence, and finally destabilizing into irregular vortex shedding at the trailing edge.
Those physical visualizations are consistent with the arguments on singing features suggested above in spectra analysis.

\subsection{Flexible blade without cavitation}
\label{Fbwithc}

\begin{figure}
\centering
\centering\includegraphics[width=0.9\linewidth]{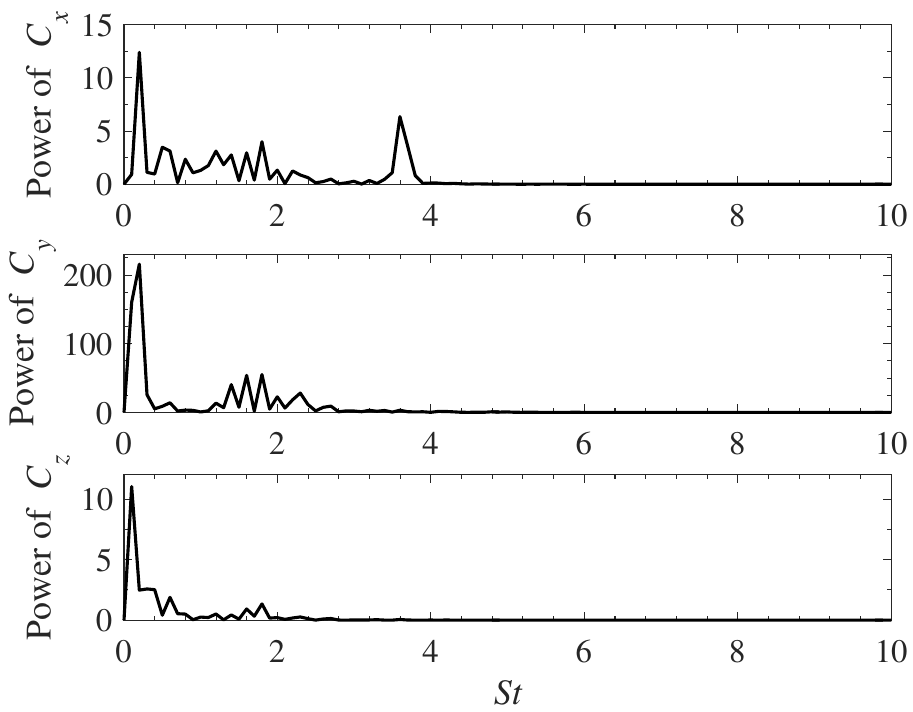}
\caption{Spectrum of lift coefficients in $x-$, $y-$, and $z-$directions for flexible hydrofoil without cavitation.}
\label{noca_9degU10_0.2E_Cxyz_fft}
\end{figure}

To follow up on the above analysis, we further analyze the potential impact of FIV on noise sources in this section.
The same hydrofoil is determined as the flexible cantilever pattern, with Young's Module $E$ of 6 $\times$ $10^{8}$ $Pa$, Poisson's ratio $\nu$ of 0.33, and mass ratio $m^*$ of 7.65.
The time histories of force coefficients $C_{x,y,z}$ and the corresponding spectrum of correlated fluctuating components are exhibited in Figs. \ref{noca_9degU10_0.2E_Cxyz_timehistory} and \ref{noca_9degU10_0.2E_Cxyz_fft}.
In addition to the broadband components, a perusal of the spectrum (cf. with Fig. \ref{noca_9degU10_0.2E_Cxyz_fft}) indicates two distinctive features: The components corresponding to $St$ = 0.2 becomes stronger for all the spectrum of $C_x$, $C_y$ and $C_z$; One more intense peak of $St$ = 3.60 appears on the spectrum of $C_x$.

\begin{figure}
\centering
\centering\includegraphics[width=1.0\linewidth]{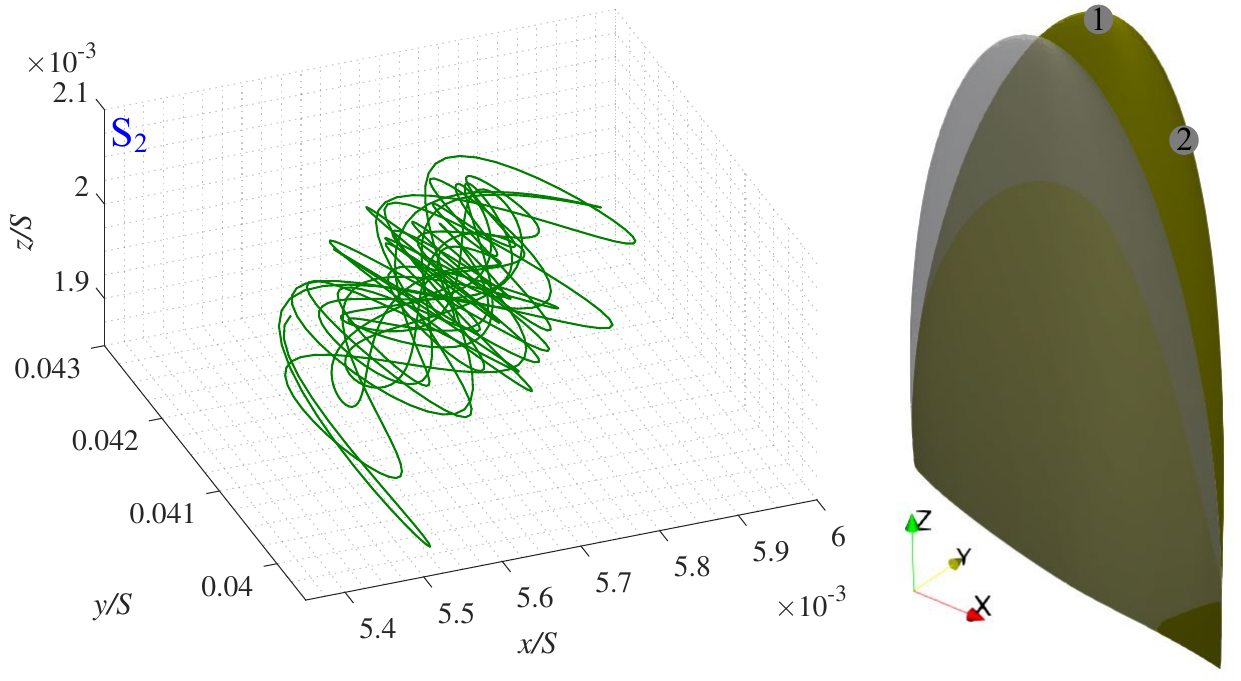}
\caption{3D trajectory of surface monitor $S_2$ for flexible hydrofoil without cavitation.}
\label{noca_9degU10_0.2E_S2_xyz}
\end{figure}

\begin{figure}
\centering
\centering\includegraphics[width=1.0\linewidth]{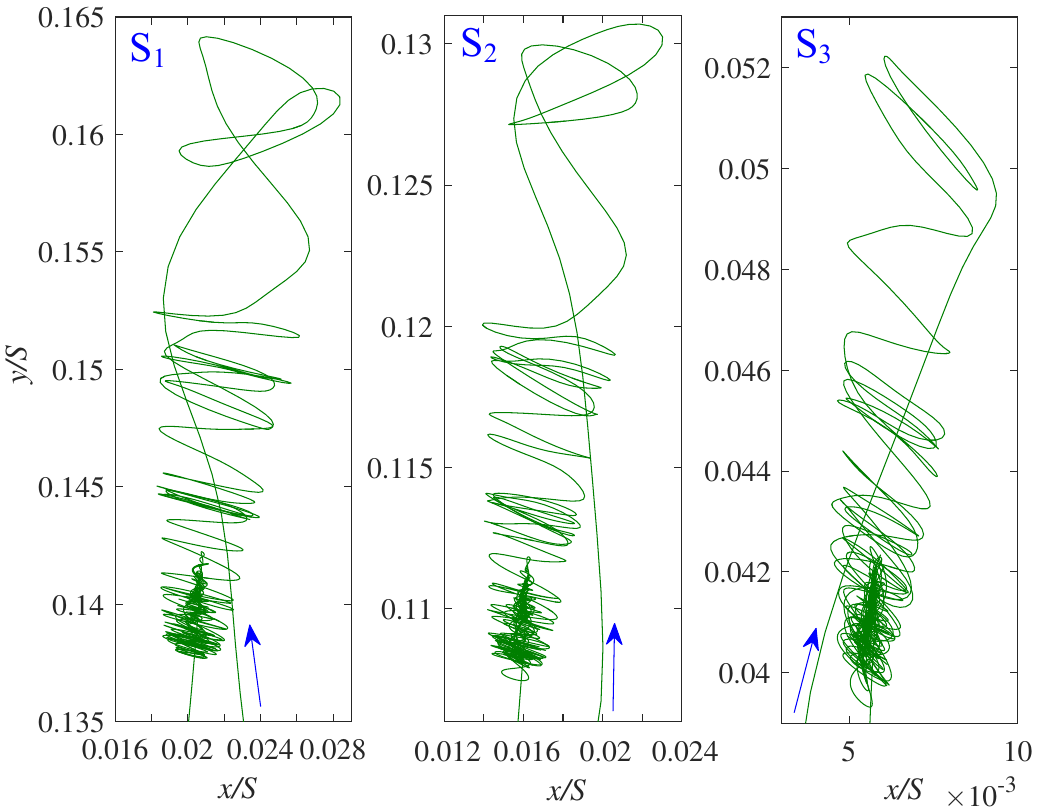}
\caption{Trajectories of the three surface monitoring points in the $x-y$ plane.}
\label{noca_9degU10_0.2E_S123_xy}
\end{figure}

The motion curve of  3D displacement at surface point \#2 (i.e., S$_2$) is shown in Fig. \ref{noca_9degU10_0.2E_S2_xyz}, and it is found that the amplitude in $z-$direction is relatively small. Therefore, we continue to depict the trajectories in the $x-$ and $y-$directions of three monitors on the surface in Fig. \ref{noca_9degU10_0.2E_S123_xy}, with the blue arrows representing the directions of the displacements. 
The displacement amplitude range of the three monitors decreases as the spanwise height (i.e., $z-$coordinate) becomes smaller.
This behavior is consistent with the instantaneous geometric illustration of structural deformation shown in Figure \ref{noca_9degU10_0.2E_S2_xyz}.

The first feature is owing to the hydrofoil being subjected to forced-vibration in the $y-$direction. Specifically, the vibration frequency is still dominated by the energetic periodic force brought about by the tip and trailing-edge vortex shedding. 
As for the structural response in the $x-$direction, it could be observed from Fig. \ref{Cxyz_timehistory} as well as Fig. \ref{Cxyz_fft}a that the fluctuation energy of the lift force in  $x-$direction is relatively tiny for the rigid structure, hence it is difficult to have a substantial effect on the structural deformation. Therefore, the structural natural frequency corresponding to hydrofoil mode \#4 (with vibration in the $x-$direction) appears in the spectrum of $C_x$ (cf. with Fig. \ref{noca_9degU10_0.2E_Cxyz_fft}a).
According to the noise analysis of cylinder vibration \cite{CHENG2023652}, it is known that the vibration of a structure in a certain direction will also intensify/amplify the noise level in that direction.

\begin{figure}
\centering
\centering\includegraphics[width=0.8\linewidth]{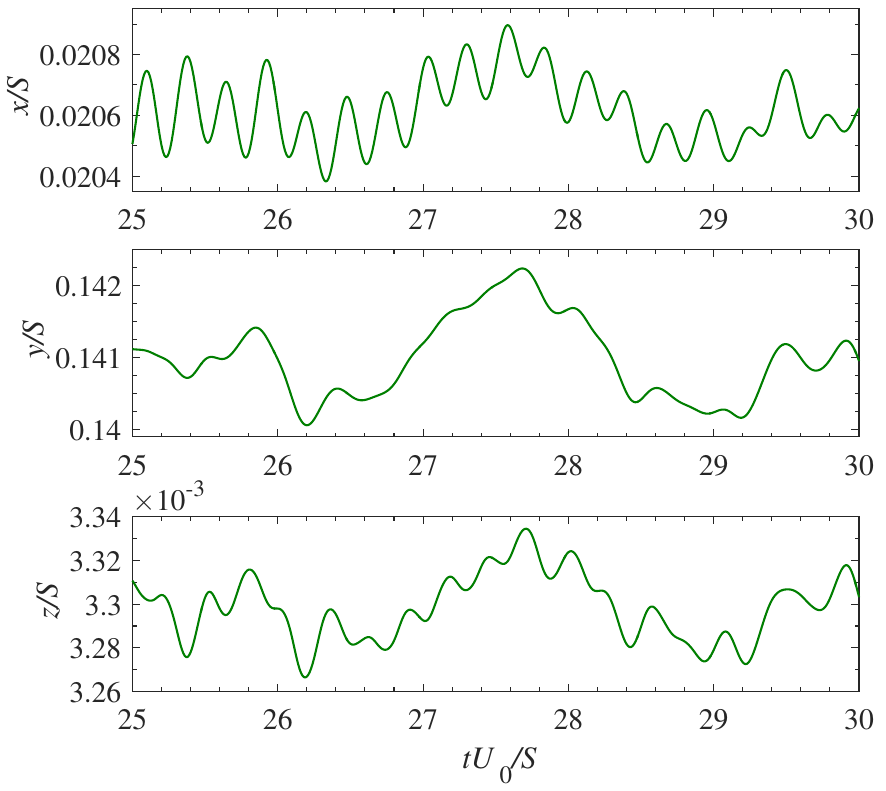}
\caption{Time-histories of structural displacements in $x-$, $y-$, and $z-$directions at surface monitor $S_1$ for flexible hydrofoil without cavitation.}
\label{noca_9degU10_0.2E_S1_xyz_time}
\end{figure}

\begin{figure}
\centering
\centering\includegraphics[width=0.75\linewidth]{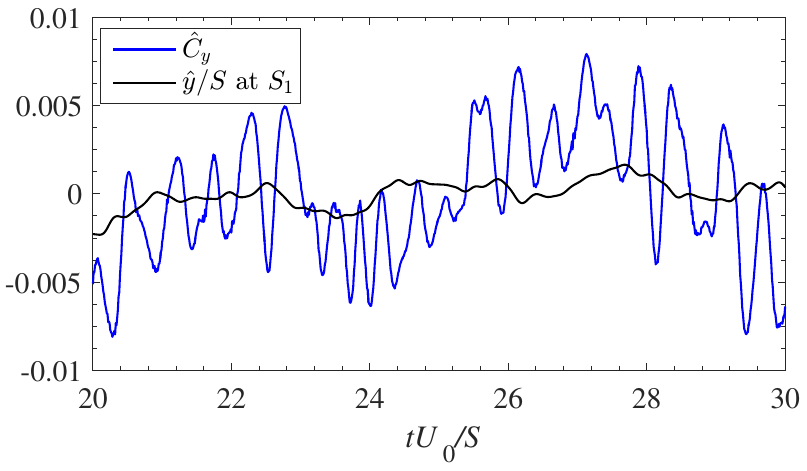}
\caption{Time-histories of fluctuating components $\hat{C}_y$ and $\hat{y}/S$ in lift coefficients $C_y$ and normalized displacements $y/S$, respectively, at surface monitor $S_1$ for flexible hydrofoil without cavitation.}
\label{Cy_yS1_fcomponents_Fnoca}
\end{figure}

\begin{figure}[h]
\centering
\centering\includegraphics[width=0.9\linewidth]{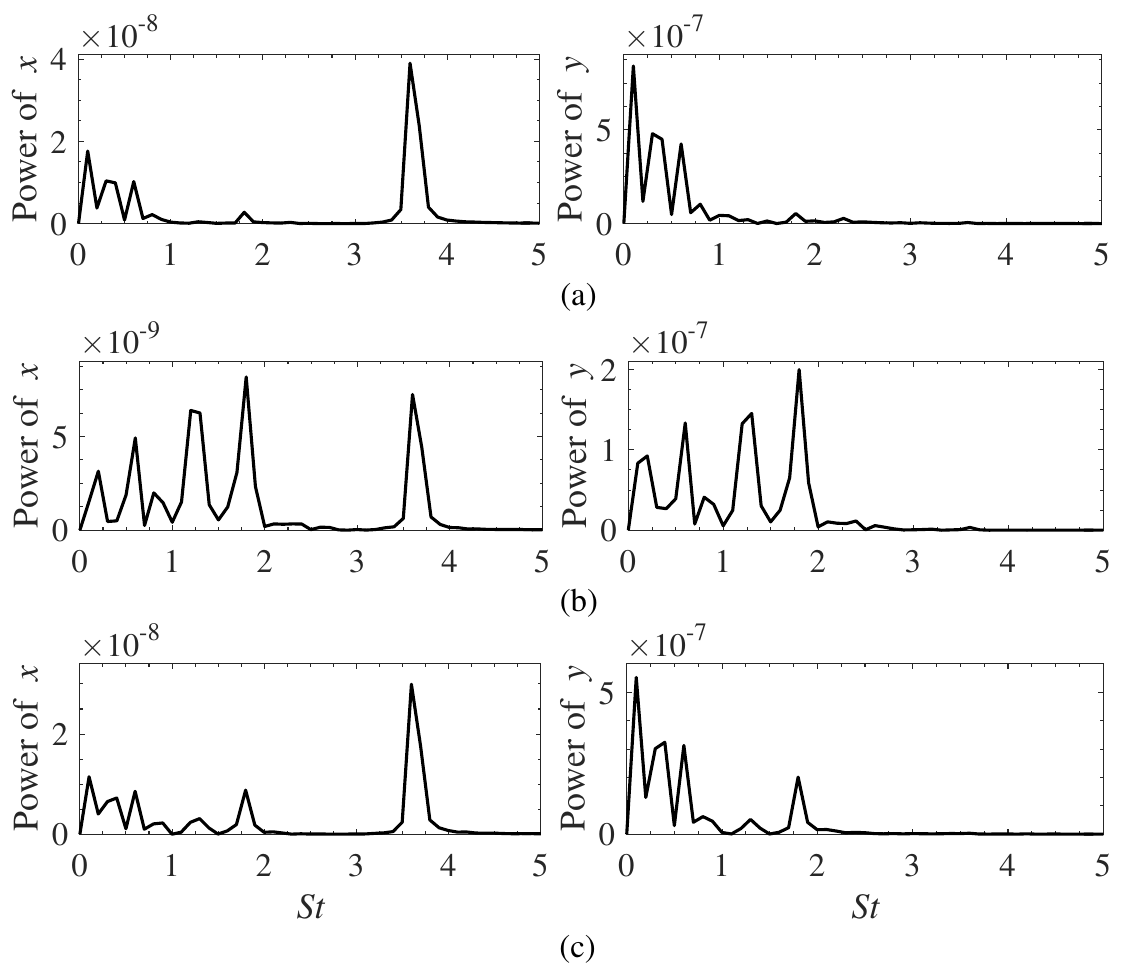}
\caption{Spectrum of structural displacements in $x-$ and $y-$directions at the location of three surface monitors for flexible hydrofoil without cavitation.}
\label{noca_9degU10_0.2E_S123_xy_FFT}
\end{figure}

\begin{figure}[h]
\centering
\centering\includegraphics[width=1.0\linewidth]{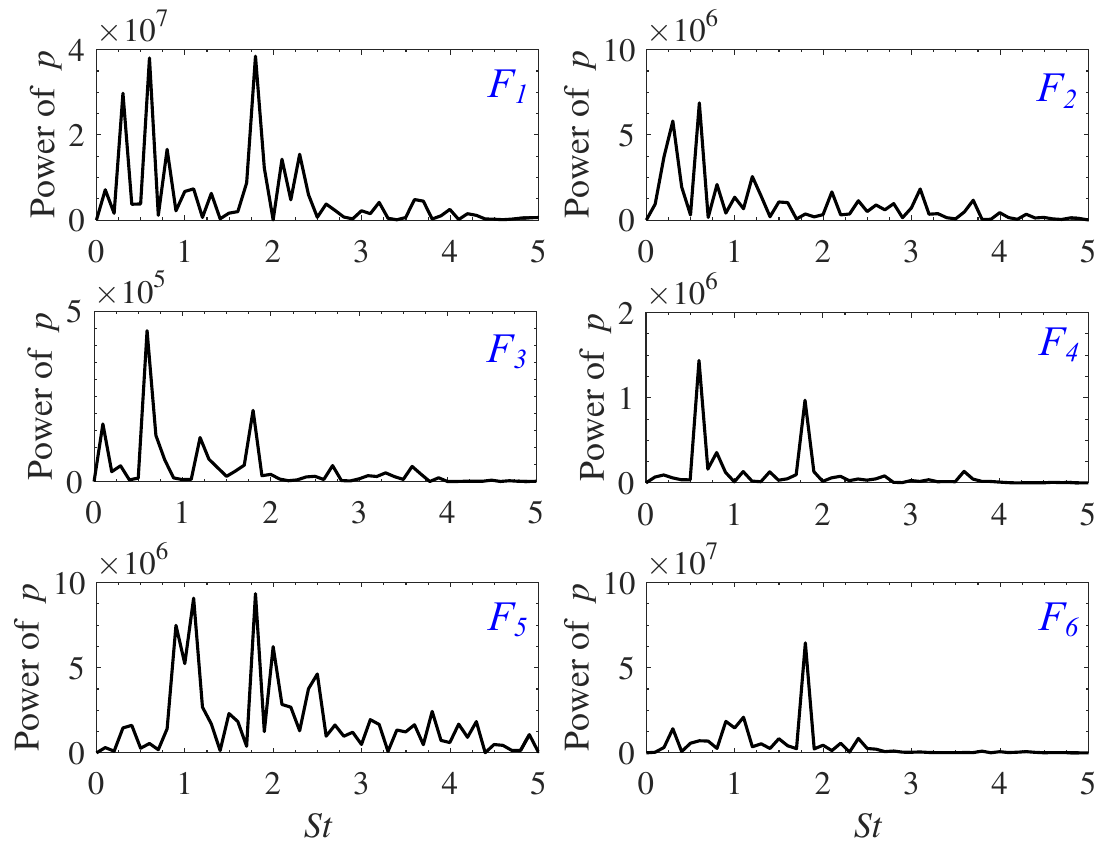}
\caption{Spectrum of pressure fluctuation at monitors $F_{1,2,3,4,5,6}$ for flexible hydrofoil without cavitation.}
\label{noca_9degU10_0.2E_P123489_fft}
\end{figure}

Figure \ref{noca_9degU10_0.2E_S1_xyz_time} displays the time-histories of structural displacements of monitor $S_2$ in $x-$, $y-$, and $z-$directions. 
The fluctuating components inside lift coefficients $C_y$ and normalized displacements $y/S$ (at $S_1$) are both displayed inside Figure \ref{Cy_yS1_fcomponents_Fnoca}. It is observed that the fluctuation of the lift coefficients on large scales is dominated by the structural vibrations, while the structure is also in turn affected by the high-frequency components of the lift.
Furthermore, the spectrum of the structural displacements in the $x-$ and $y-$directions for the three monitoring points $S_{1,2,3}$ are depicted in Fig. \ref{noca_9degU10_0.2E_S123_xy_FFT}.
The dominating peak frequency of $x-$displacements is located at $S_t$ of 3.60, consistent with that of force coefficient $C_x$. Moreover, one additional intense peak with $St$ of 1.80 appears in the displacements spectrum of both $x-$ and $y-$directions.
However, from Fig. \ref{noca_9degU10_0.2E_P123489_fft} it is clear that this spectrum component does not originate from the natural frequency of any of the structural modes.
In this case, it is suggested that this frequency component with $St$ of 1.80 originates from the effect of the periodical force owing to the vortex shedding from the hydrofoil surface. More specifically, the deformation of the hydrofoil affects the mechanism of trailing-edge vortex-shedding structure and induces the frequency components with $St$ of 1.80 inside surface forces (cf. with Figure \ref{noca_9degU10_0.2E_Cxyz_fft} a,b). This thereby has a considerable impact on structural deformation and motion.

For the propeller singing problem, the vortex-shedding behaviors in the wake directly affect the magnitude and frequency of the noise source. To further explore this, we extracted the pressure variation at several points in the flow field and displayed their spectra in Fig. \ref{noca_9degU10_0.2E_P123489_fft}. 
Four monitors, $F_{1, 2, 3, 4}$, are located in vicinity of the tip vortex trajectory, while the other two points,  $F_{5,6}$, are close to the trailing edge.
To begin, it can be observed that the pressure fluctuations near the tip vortex still contain the dominant frequency of $St$ = 0.2. Moreover, pressure at all monitors also contains the dominant frequency component, i.e., the peak corresponding to $St$ = 1.80. 
Since the pressure fluctuations on the trajectory of the tip vortex are not directly affected by the trailing-edge vortex shedding, it is expected that they come from the effect of the blade vibration. 
In this case, we infer that the overall process should perform as: deformation of the hydrofoil leads to a change in the form of the trailing-edge vortex shedding, and also the appearance of a new component of surface force corresponding to $St$ = 1.80. 
This component has a in turn implication on the hydrofoil vibration. The hydrofoil tip vibration then transfers this frequency component to the pressure fluctuations in the tip vortex, which thereby affects the noise source. 
Therefore, it is suggested that the vibration of the hydrofoil structure will be closely correlated to the characteristics of the propeller URN and also singing behaviors.

\begin{figure}[h]
\centering
\centering\includegraphics[width=1.0\linewidth]{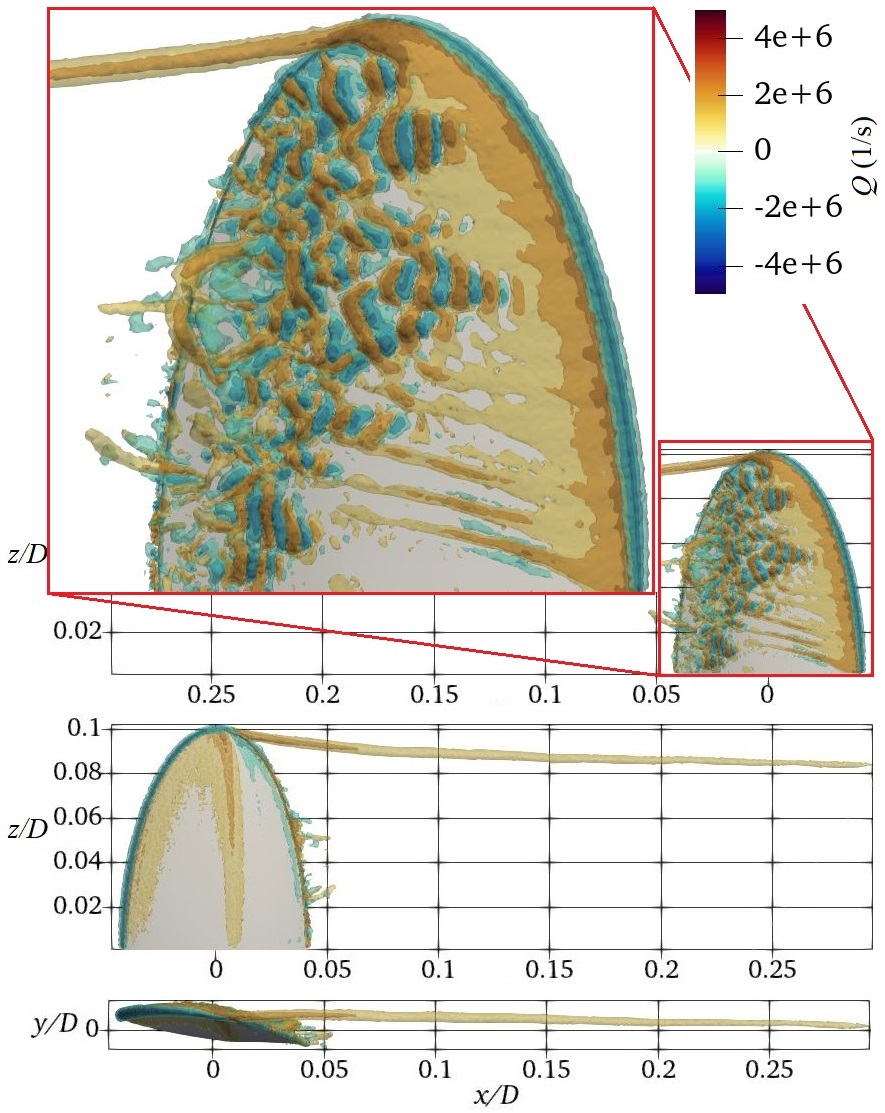}
\caption{Instantaneous isopleths of the second invariant $Q$ from different overviews and enlarged view on the suction side for flexible hydrofoil without cavitation.}
\label{noca_E04_Q_3D}
\end{figure}

\begin{figure}[h]
\centering
\centering\includegraphics[width=0.8\linewidth]{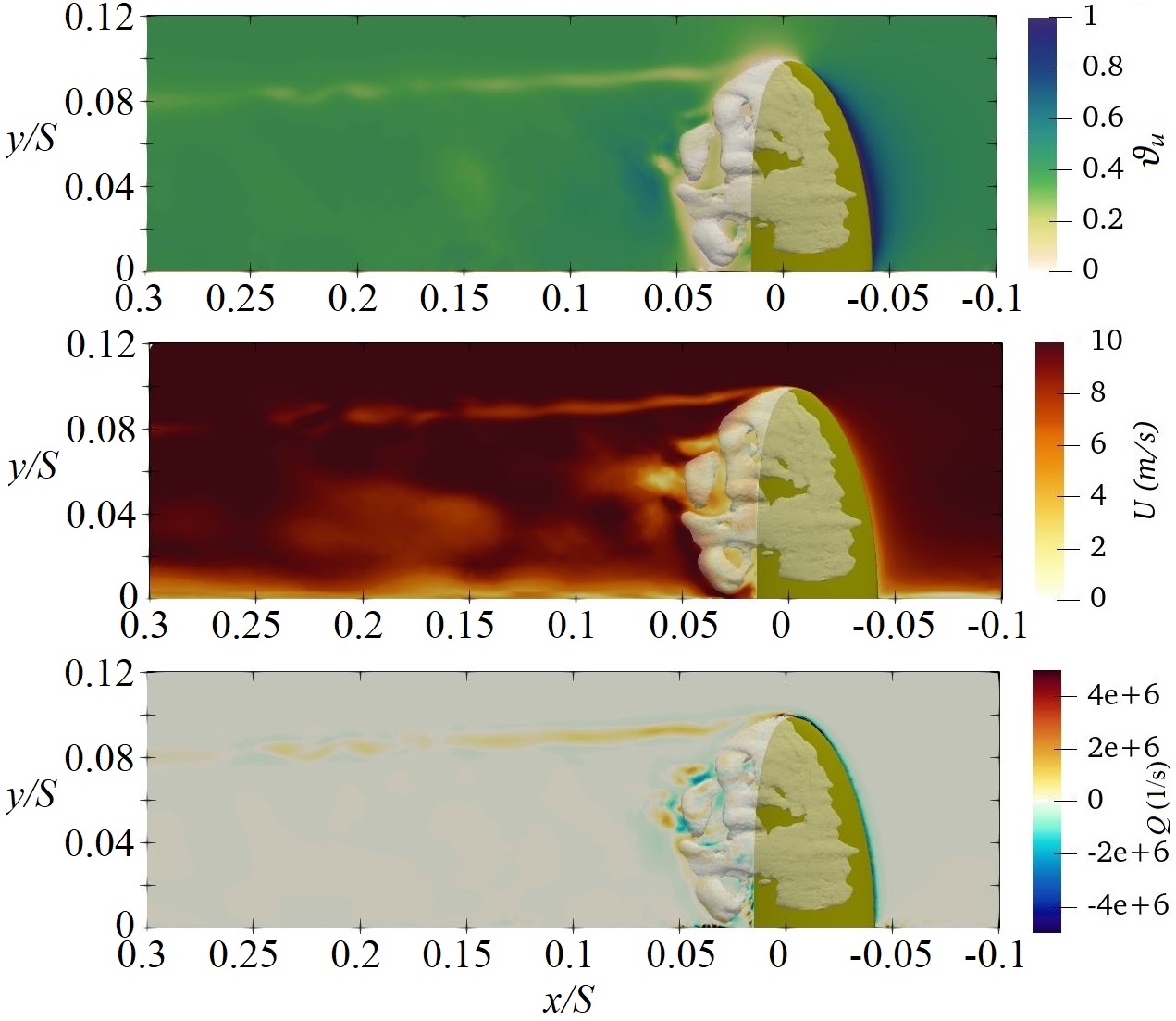}
\caption{Cavitation contour accompanied by the $x-y$ planes, on which cavitation number $\vartheta_n$, velocity $U$, and second invariant of the velocity gradient tensor $Q$ are depicted in panels 1, 2, and 3, respectively.}
\label{ca_E04_cavitation_slice_xz0}
\end{figure}

Figure \ref{noca_E04_Q_3D} exhibits the second invariant $Q$ of the velocity gradient tensor under a Galilean transformation. 
It should be noted that $Q$ represents the difference between rotation rate and strain rate.
Its positive value indicates that the vortex structure is rapidly forming at a correlated location, and negative value means that the vortex structure is being dissipated.
As can be seen from the top view, there is a significant deformation of the hydrofoil, and there is a clear vortex shedding behind the blade tip.
In addition, compared with the pressure side of the blade, there are a large number of areas with positive and negative values of $Q$ on the suction side of the blade, indicating abundant vortex structures on the suction side. 
Furthermore, the structural vibration will inevitably affect the vortex generation and shedding at the trailing-edge as well as at the tip, which in turn will have an impact on the noise sources in the flow field. This is also demonstrated in the above mathematical analysis.

\subsection{Flexible blade with cavitation}

To investigate the impact of cavitation generation on structural response and noise sources, in this section, the Schnerr-Sauer cavitation model is added in the calculation of fluid-structure interaction for the present flexible hydrofoil compared to the setting in subsection \ref{Fbwithc} and other properties maintain constant. 
Figure \ref{ca_E04_cavitation_slice_xz0} exhibits the cavitation contour (with volume fraction of 0.5) accompanied by the $x-y$ planes, on which cavitation number $\vartheta_n$, velocity $U$, and second invariant of the velocity gradient tensor $Q$ are depicted, as shown in panels 1, 2, and 3, respectively.
Firstly, a large amount of sheet cavitation occurs in present numerical results, while no tip vortex cavitation occurs, even though all dynamical information in the three panels indicates the presence of a tip vortex. 
It is expected that the disappearance of tip vortex cavitation is because the mesh quality in this calculation may not be sufficient to resolve the cavitation structure, although it could accurately estimate the flow field and surface pressure. Further work is needed in the future to achieve the capture of the tip vortex cavitation using a higher accuracy mesh strategy. Particular emphasis of the present work lies on the effect of sheet cavitation generation on the structural response and noise sources.

\begin{figure}
\centering
\centering\includegraphics[width=0.8\linewidth]{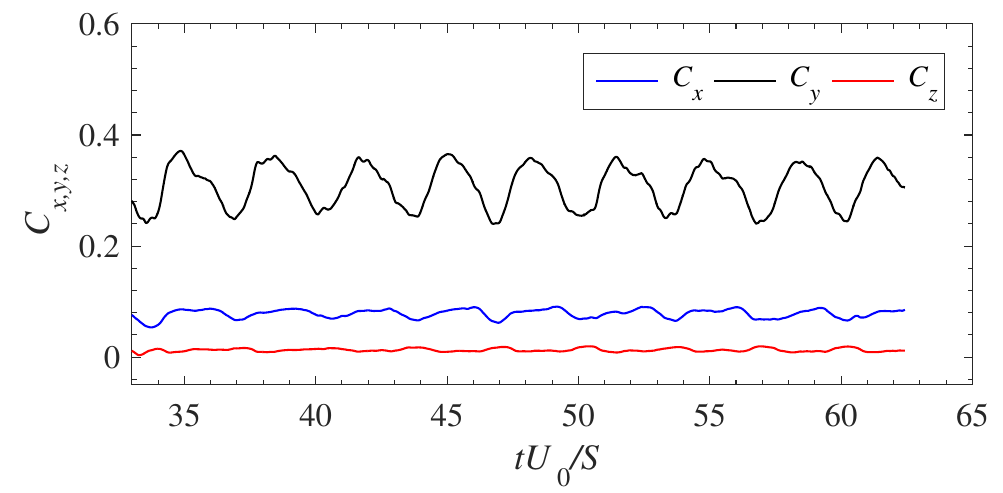}
\caption{Time-histories of lift coefficients in $x-$, $y-$, and $z-$directions for flexible hydrofoil with sheet cavitation.}
\label{ca_9degU10_E04_Cxyz_t}
\end{figure}

\begin{figure}[h]
\centering
\centering\includegraphics[width=0.85\linewidth]{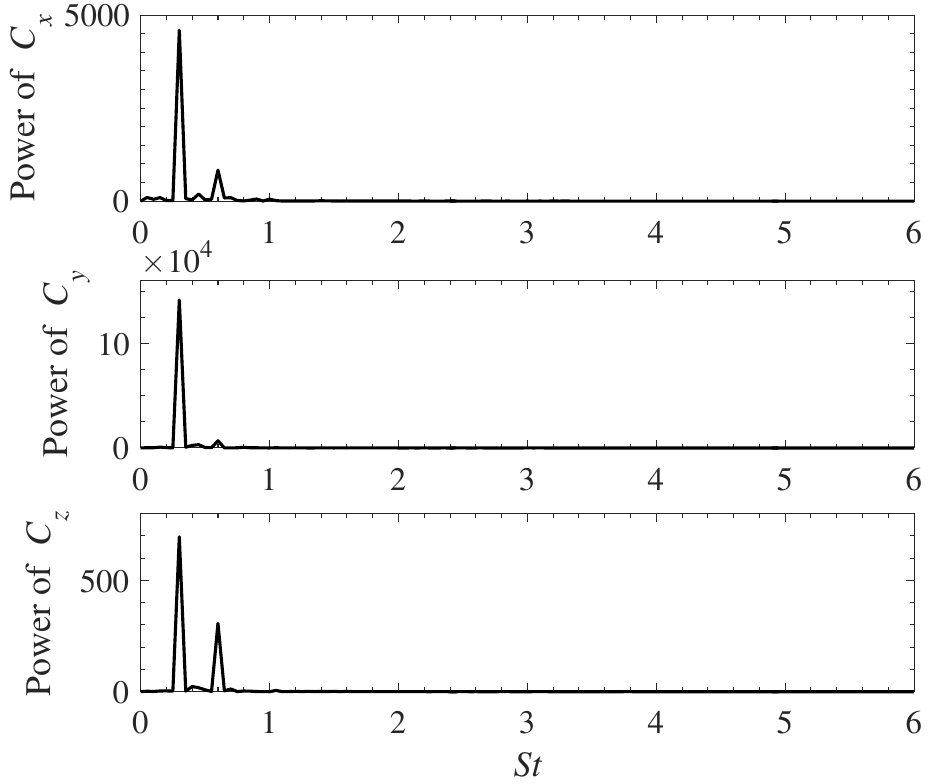}
\caption{Spectrum of lift coefficients in $x-$, $y-$, and $z-$directions for flexible hydrofoil with sheet cavitation.}
\label{ca_9degU10_E04_Cxyz_fft}
\end{figure}

\begin{figure}
\centering
\centering\includegraphics[width=0.85\linewidth]{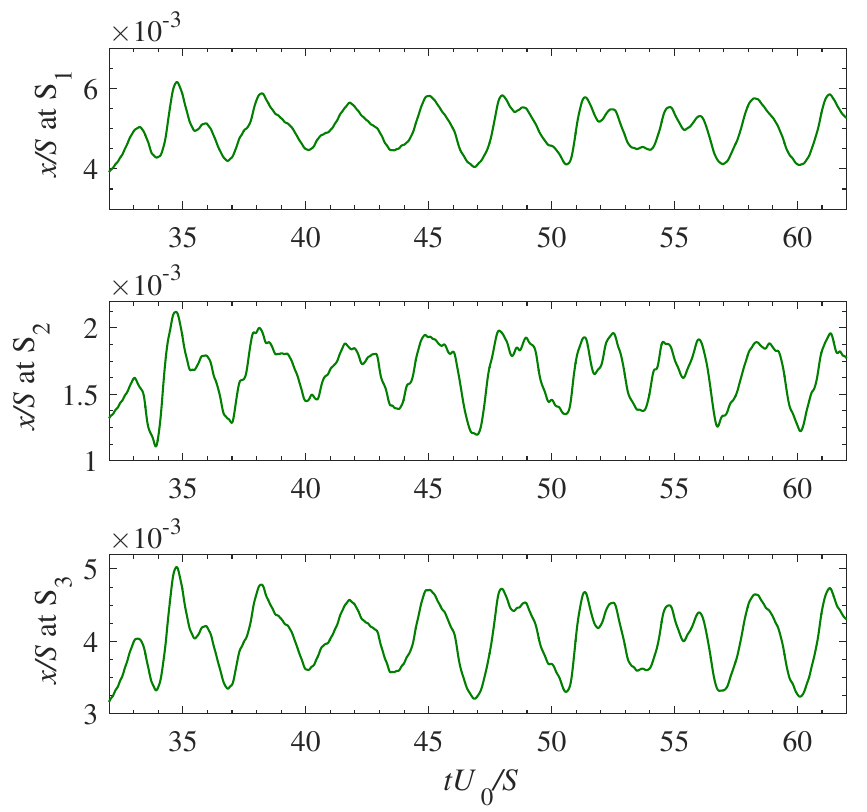}
\caption{Time-histories of structural displacements in $x-$direction at three surface monitors for flexible hydrofoil  with sheet cavitation.}
\label{ca_9degU10_0.4E_S123_x_t}
\end{figure}

\begin{figure}
\centering
\centering\includegraphics[width=0.85\linewidth]{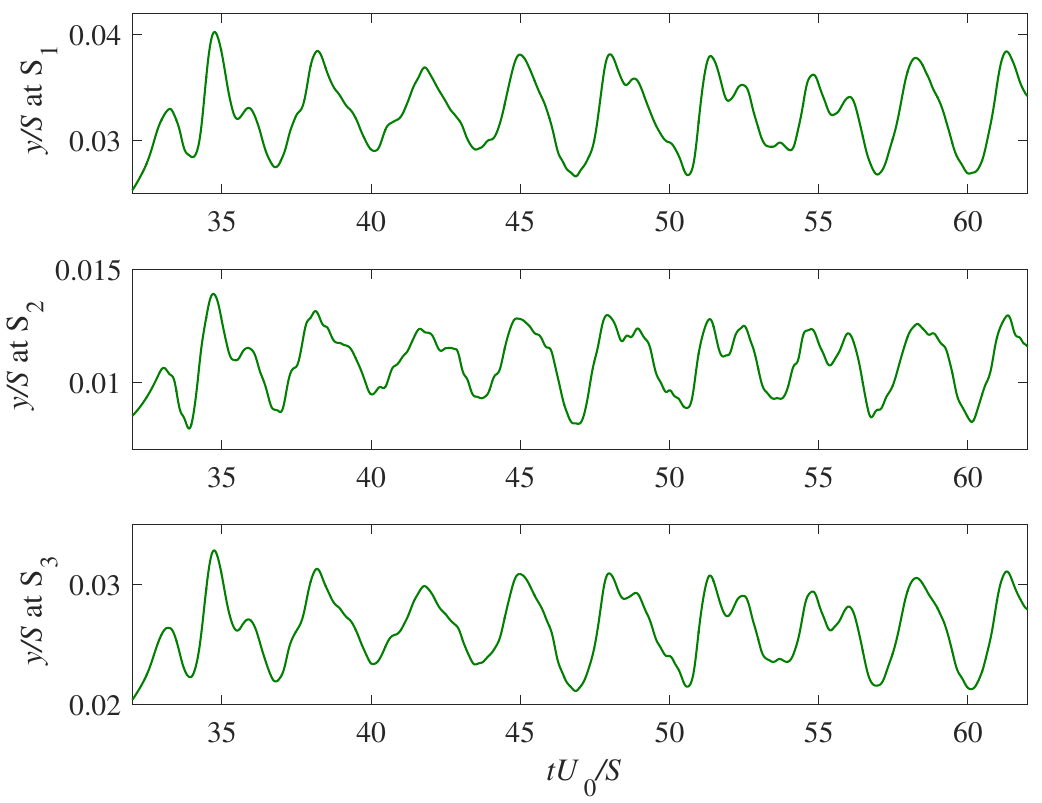}
\caption{Time-histories of structural displacements in $y-$direction at three surface monitors for flexible hydrofoil  with sheet cavitation.}
\label{ca_9degU10_0.4E_S123_y_t}
\end{figure}

The time-histories and corresponding spectrum of the lift coefficients in three directions are displayed in Figures \ref{ca_9degU10_E04_Cxyz_t} and \ref{ca_9degU10_E04_Cxyz_fft}, respectively.
The formation of sheet cavitation leads to a significant change in the flow dynamics. To begin, the shedding of sheet cavitation from the trailing edge of the blade is regular, which leads to a dominant peak (at $St$ = 0.30), and another harmonic component (at $St$ = 0.60) in the spectrum of the surface force coefficients in all three directions in Fig. \ref{ca_9degU10_E04_Cxyz_t}. In addition, comparing Fig. \ref{ca_9degU10_E04_Cxyz_fft} with Fig. \ref{noca_9degU10_0.2E_Cxyz_timehistory}, the generation of cavitation also leads to a significant increase in the energy of the fluctuating components in surface forces, which will result into amplified structural vibrations.

\begin{figure}
\centering
\centering\includegraphics[width=0.8\linewidth]{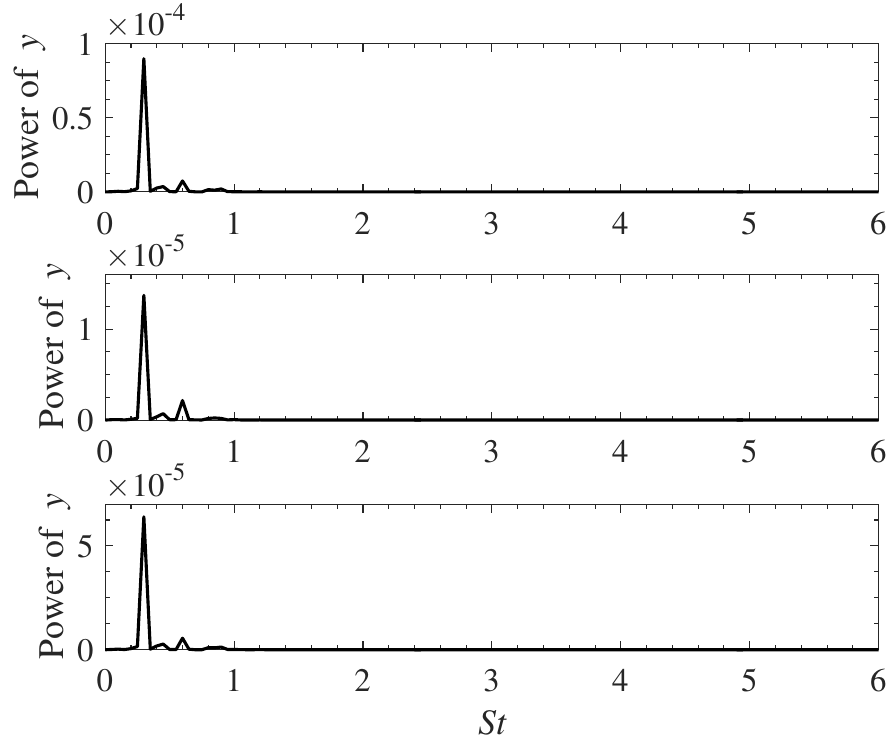}
\caption{Spectrum of structural displacements in $y-$direction at three surface monitors for flexible hydrofoil with sheet cavitation.}
\label{ca_9degU10_E04_S123_y_fft}
\end{figure}

\begin{figure}
\centering
\centering\includegraphics[width=0.95\linewidth]{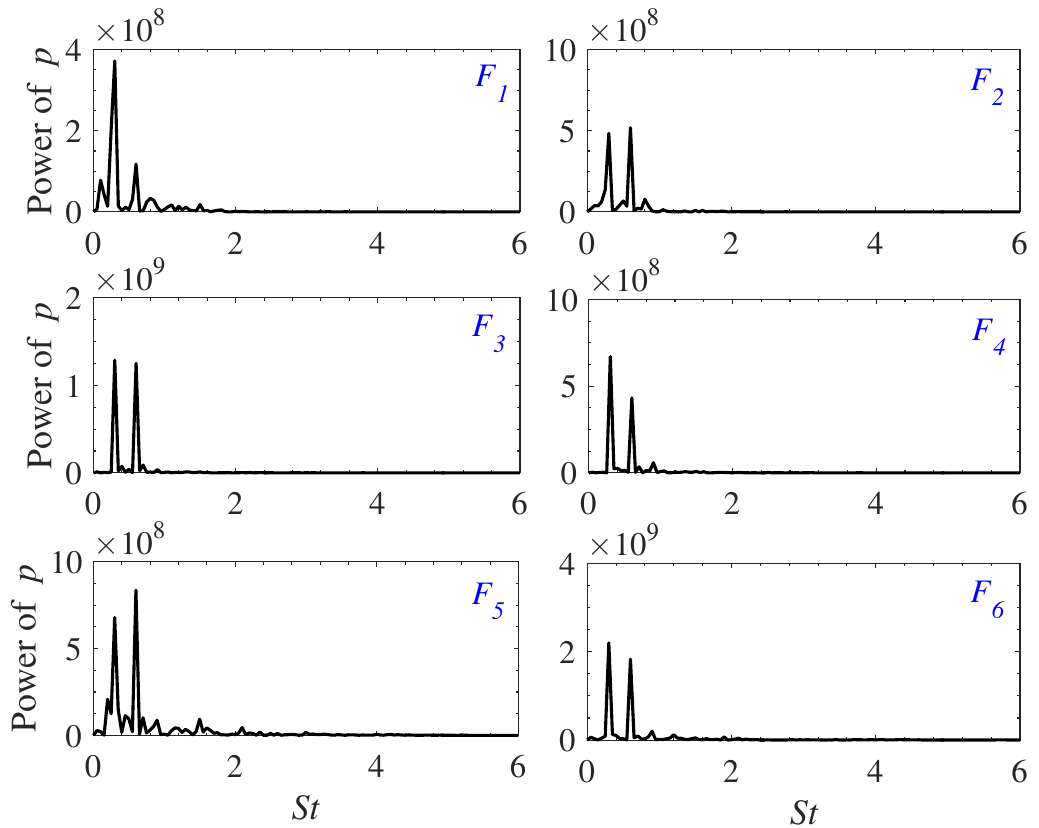}
\caption{Spectrum of pressure fluctuation at monitors $F_{1,2,3,4,5,6}$ for flexible hydrofoil with sheet cavitation.}
\label{ca_9degU10_E04_p123489_fft}
\end{figure}

In this case, the structural displacements in $x-$ and $y-$directions for three surface monitors are displayed in Figures. \ref{ca_9degU10_0.4E_S123_x_t} and \ref{ca_9degU10_0.4E_S123_y_t}, respectively.
A comparison of Fig. \ref{noca_9degU10_0.2E_S1_xyz_time} and Fig. \ref{ca_9degU10_0.4E_S123_x_t} indicates that the structural vibration amplitudes are significantly increased owing to the sheet cavitation.
Additionally, the overall structural response implies the feature of flutter behavior, where the hydrofoil structural response resonates with variations of lift and their harmonics herein.
This is demonstrated by Fig. \ref{ca_9degU10_E04_S123_y_fft}, in which the transverse ($y-$) displacement spectrum includes identical peaks as that in Fig. \ref{ca_9degU10_E04_Cxyz_fft}.
The spectrum of the pressure at monitors $F_{1,2,3,4,5,6}$ in the flow fields are plotted in Fig. \ref{ca_9degU10_E04_p123489_fft}.
Compared to the non-cavitating situation in Fig .\ref{noca_9degU10_0.2E_S123_xy_FFT}, the intense peaks of the spectrum at all monitoring points are concentrated at $St$ = 0.3 and 0.6, except for the pressure spectrum at $S_1$ (whose location is immediately adjacent to the blade tip), where a peak at around $St$ = 0.2 still occurs.
In conjunction with the above discussion, such structural vibration necessarily affects the pressure fluctuations within the flow field, which in turn affects the characterization of the noise source. 
Hence, it could be concluded here the cavitation and structural vibrations co-dominate the generation of noise sources for propeller singing behaviors. 
Moreover, there is also a reciprocal effect between cavitation structures and hydrofoil vibrations.
If tip vortex cavitation appears, it would also be influenced by structural vibration and thereby could impact the propeller singing behavior.

\section{Conclusion}
Using the recently developed cavitation FSI solver, this paper studied the coupled dynamics of flow-induced vibration and cavitation of cantilever flexible hydrofoils and explored the potential impact on hydro-noise sources. The present work involved the computation of three cases: a rigid hydrofoil without cavitation, a flexible hydrofoil without cavitation, and a flexible hydrofoil with cavitation. The comparative analysis indicated that the pressure fluctuations within the flow field are mainly affected by the tip vortex shedding, the trailing-edge vortex shedding, and the structural vibration. Among them, the tip vortex shedding and blade vibration are correlated to the intense peak (low-frequency tonal components) of the noise source, while the trailing-edge vortex shedding leads to broadband noise. We further analyzed the effect of sheet cavitation's appearance on the flow-solid interaction response. The results demonstrated that the shedding of sheet cavitation induces considerable periodic forces on the hydrofoil surface, which induces flutter-like response, dominates the blade vibration, and affects the pressure fluctuations in the flow field. The frequency of tonal compoments inside potential noise sources are consistent with those of cavitation shedding behaviors and structural vibration.
It could be summarized that the cavitating behaviors and structural vibrations co-dominate the features of noise sources for propeller singing behavior. 
Moreover, there is also a reciprocal effect between cavitation structures and hydrofoil vibrations. The deformation and oscillation of hydrofoil will also in turn modify the vortex and cavitation shedding dynamics.
In the future, further refined meshing strategy will be considered to resolve the tip vortex cavitation and examine the underlying mechanism of noise generation and propagation. Additionally, detailed work should be conducted to locate the resonance range of the considered FIV system.

\section*{Acknowledgment} 
The present study is supported by Mitacs, Transport Canada and Clear Seas through a Quiet-Vessel Initiative (QVI) program. This work was made possible by the facilities of the Shared Hierarchical Academic Research Computing  (\href{http://www.sharcnet.ca}{SHARCNET}) and Compute/Calcul Canada.

\bibliographystyle{asmeconf}  
\bibliography{Psing_OMAE}

\appendix   



\selectlanguage{english} 

\end{document}